\documentclass[a4paper,final]{aa} 

\usepackage{epsf,epsfig,graphicx,amsmath}

  \def\bruch#1#2{\frac{#1}{#2}}
  \def\abl#1#2{\bruch {d #1}{d #2}}
  \def\pabl#1#2{\bruch {\partial #1}{\partial #2}}
  \def\Vl{{V_{\ell}}}
  \def\rr{\vec{r}}
  \def\nn{\vec{n}}
  \def\er{\vec{e}_r}
  \def\vv{\vec{v}}
  \def\Tg{T_{\rm g}}
  \def\Td{T_{\rm d}}
  \def\Trad{T_{\rm rad}}
  \def\Tref{T_{\rm ref}}
  \def\kext{\widehat{\kappa}_{\rm ext}}
  \def\nH{n_{\langle{\rm H}\rangle}}
  \def\tauL{\tau_{\rm L}}
  \def\tauref{\tau_{\rm ref}}
  \def\Rst0{R_\star^{\,0}}
  \def\etal{${\rm \hspace*{0.8ex}et\hspace*{0.7ex}al.\hspace*{0.65ex}}$}
  \def\plus{${\rm \hspace*{0.7ex}\&\hspace*{0.7ex}}$}
  \def\ie{i.\,e.\ }
  \def\eg{e.\,g.\ }

\begin{document}

\title{2D Models for Dust-driven AGB Star Winds}

\author{Peter Woitke}

%\offprints{P.~Woitke (woitke@strw.leidenuniv.nl)}

\institute{Sterrewacht Leiden, P.O.~Box 9513, 2300 RA Leiden, The Netherlands,
           \email woitke@strw.leidenuniv.nl } 

\date{Received date; accepted date}

\abstract{New axisymmetric (2D) models for dust-driven winds of
  C-stars are presented which include hydrodynamics with radiation
  pressure on dust, equilibrium chemistry and time-dependent dust
  formation with coupled grey Monte Carlo radiative
  transfer. Considering the most simple case without stellar pulsation
  (hydrostatic inner boundary condition) these models reveal a more
  complex picture of the dust formation and wind acceleration as
  compared to earlier published spherically symmetric (1D) models. The
  so-called exterior $\kappa$-mechanism causes radial oscillations
  with short phases of active dust formation between longer phases
  without appreciable dust formation, just like in the 1D models.
  However, in 2D geometry, the oscillations can be out-of-phase at
  different places above the stellar atmosphere which result in the
  formation of dust arcs or smaller caps that only occupy a certain
  fraction of the total solid angle. These dust structures are
  accelerated outward by radiation pressure, expanding radially and
  tangentially like mushroom clouds, while dust-poor matter is falling
  back towards the star at other places.  A highly dynamical and turbulent
  dust formation zone is created in this way, which again leads to
  inhomogeneous dust production. Further away from the star, flow
  instabilities (\eg Rayleigh-Taylor) have time to fragment the
  outward moving arcs and shells to produce numerous small-scale 
  cloud-like sub-structures.

  }

\maketitle
\sloppy

%----------------------------------------------------------------------------
\section{Introduction}
%----------------------------------------------------------------------------
Asymptotic Giant Branch (AGB) stars are known to generate massive
stellar winds with mass loss rates up to $10^{-4}\,\rm
M_\odot/yr$ (Habing 1996, LeBertre 1997, Wallerstein \& Knapp
1998)\nocite{habing96,lebertre97,wk98}, which drives them towards the
planetary nebula phase. According to the currently available
spherically symmetric (1D) model calculations (Winters et al.\ 2000,
H{\"o}fner\etal 2003, Sandin\plus H{\"o}fner 2003, Schirr\-macher\etal
2003)\nocite{wlj2000,hga2003,sh2003,sws2003}, these winds are driven
by a combination of stellar pulsation and radiation pressure on dust
grains. An exterior $\kappa$-mechanism leads to the production of
radial dust shells in more or less regular long time intervals, even
if the pulsation of the star is neglected in the model (Fleischer\etal
1995, H{\"o}fner\etal 1995).  \nocite{fgs95,hfd95}

The formation of radial dust shells, however, does not seem to be in
general agreement with recent interferometric observations. Infrared
speckle observations have clearly indicated that the innermost dust
distribution around red giants can deviate strongly from spherical
symmetry (Weigelt\etal 2002, Monnier\etal 2004, Tuthill\etal 2005)
with evidence for moving clouds, arcs and ``bars''. The new IR {\sc
Vlti} instruments {\sc Amber} and {\sc Midi} will allow for an even
more detailed view on the dust formation and wind acceleration zones
of AGB stars with the ability to detect wind asymmetries and
cloud-like inhomogeneities, and to track individual dust shells or
clouds.  \nocite{tmd2005,wbb2002,mmt2004}

These observations call for likewise detailed 3D models. Without such
models, the interpretation of the new interferometric data will
necessarily remain on a more or less phenomenological level which is
unsatisfactory because it leaves most of the important physical
questions unanswered.

The major aim of this work is to advance from 1D to 2D dust-driven
wind models. These models will enable us to discuss whether
dust-driven winds remain spherically symmetric (if the initial and
boundary conditions are spherically symmetric) or whether
instabilities lead to spontaneous symmetry breaking, structure
formation and more complex flow patterns, which might explain the
observed irregular dust structures. Post-processing of the model
results will allow us to calculate simulated images and visibilities
(see Woitke\plus Quirrenbach 2005) which can be directly compared to
observations. \nocite{wq2005}

%----------------------------------------------------------------------------
\section{The model}
%----------------------------------------------------------------------------

\subsection{The equations}
%=========================

The dust-driven stellar wind is described by the equation of
continuity (Eq.\,\ref{eq1}), the equation of motion (Eq.\,\ref{eq2}),
the gas energy equation (Eq.\,\ref{eq3}), a system of moment equations
describing the time-dependent formation of dust particles from the gas
phase (Eq.\,\ref{eq4}), and the radiative transfer equation
(Eq.\,\ref{eq5}):
\begin{eqnarray}
  \pabl{\rho}{t} + \nabla(\rho\vv) & = & 0 
    \label{eq1}\\
  \pabl{}{t}\big(\rho\vv\big) 
    + \nabla\big(\rho\hspace*{0.2mm}
      \vv\hspace{-0.5mm}\begin{array}{c}
      {\scriptscriptstyle \circ}\\[0.45ex]\end{array}\hspace{-0.5mm}\vv 
      + p\big)  & = & 
        \rho\left(\vec{a}_{\rm rad}  - \vec{a}_{\rm grav}\right) 
    \label{eq2}\\[0.8ex]
  \pabl{}{t}\big(\rho e\big) + \nabla\big([\rho e + p]\vv\big)  & = & 
  \rho\vv\!\cdot\!\big(\vec{a}_{\rm rad}\!-\!\vec{a}_{\rm grav}\big)
      + \rho\,Q_{\rm rad} 
    \label{eq3}\\[0.8ex]
  \pabl{}{t}\big(\rho{L_j}\big)
    + \nabla\big(\rho{L_j}\vv\big) & = & 
      \Vl^{\,j/3} J_\star 
      \,+\, \frac{j}{3}\,\chi^{\rm net}\,\rho L_{j-1} 
    \label{eq4}\\[0.8ex]  
  \nn\cdot\nabla {I_\nu(\rr,\nn)} & = & \eta_\nu\!(\rr,\nn)
      -\kappa^{\rm\,ext}_\nu(\rr)\,{I_\nu}(\rr,\nn)
    \label{eq5}     
\end{eqnarray}
$\rho$ is the mass density, $e=e_{\rm int}+e_{\rm kin}$ the internal
(including kinetic) gas energy and $p$ the thermal gas
pressure. $\vec{a}_{\rm grav}$ is the gravitational deceleration
and $\vec{a}_{\rm rad}$ the acceleration by radiation pressure on
dust and gas. $Q_{\rm rad}$ is the net energy exchange rate of the gas
with the radiation field (negative for radiative cooling). For further
explanations of the introduced quantities see Sects.~\ref{sec:dust} 
and \ref{sec:radtrans}. 

The dust/gas mixture is approximated by a single hydrodynamic fluid
with bulk velocity $\vv$ (neglecting drift velocities\footnote{See
Simis\etal (2001) and Sandin\plus H{\"o}fner (2003) for the effects of
dust drift velocities in 1D models.}). However, since dust and gas are
known to decouple thermally from each other easily under the
predominant conditions in red giant winds (\eg Woitke\etal 1999), two
distinct temperatures are introduced, namely the gas temperature $\Tg$
and the dust temperature $\Td$. Whereas $\Tg$ is a result of the
solution of the gas energy equation (Eq.\,\ref{eq3}), we assume that
the dust temperature $\Td$ relaxes instantaneously to its local
radiative equilibrium value, which is part of the results of our
radiative transfer treatment (see Sect.\,\ref{sec:radtrans}).
\nocite{woi99}

\subsection{Hydrodynamics}
%=========================
\label{sec:hydro}

The multi-dimensional models are developed in the frame of the {\sc
Flash}\,2.4 hydrocode (Fryxell\etal 2000)\nocite{for2000}, which is an
explicit, finite volume, high-order Godunov-type hydro-solver which
uses Adaptive Mesh Refinement (AMR) and is fully parallel
(MPI). Spherical coordinates $(r,\theta,\phi)$ are used either in 1D
(spherically symmetric models) or 2D (axisymmetric models).

The source terms appearing on the r.h.s. of the hydro-equations
(Eqs.\,1 to 4) are treated by operator splitting, where the
time-integration is carried out by means of the implicit extrapolation
solver {\sc Limex}\,4.2A1 (Deuflhard \& Nowak 1987).\nocite{dn87}

Gravity and radiation pressure are calculated by
\begin{equation}
  \vec{a}_{\rm rad} - \vec{a}_{\rm grav}
  = \left(\frac{\kext L_\star}{4\pi r^2 c} - \frac{G M_r}{r^2}\right)\er\ ,
  \label{eq:alpha}
\end{equation}
where $\er$ is the unit vector in the radial direction, $G$ the
constant of gravity, $L_\star$ the stellar luminosity, $\kext$ the
grey (dust\,+\,gas) extinction coefficient per mass and $c$ the speed
of light. The enclosed mass is approximated by
$M_r\!\approx\!M_r(t\!=\!0)$, \ie we assume a constant spherically
symmetric gravitational potential as calculated for the initial
state. The radiation pressure is assumed to be strictly directed into the
radial direction.

Net radiative heating of the gas is calculated by
\begin{equation}
  Q_{\rm rad} = 4 \sigma \widehat{\kappa}^{\rm\,gas}_{\rm abs} 
                \big(\Trad^4-\Tg^4\big)\ ,   
\end{equation}
assuming LTE, where $\widehat{\kappa}^{\rm\,gas}_{\rm abs}$ is the gas
absorption coefficient per mass and $\Trad$ is a measure for the mean
intensity ($J=\frac{\sigma}{\pi}\Trad^4$, see
Sect.\,\ref{sec:radtrans}).

As equation of state, we apply the ideal gas law for atomic hydrogen
with $\gamma=5/3$ (disregarding H$_2$), \ie $p =
\frac{k}{\bar{\mu}}\,\rho\,\Tg$ and $e_{\rm int} =
\frac{1}{\gamma-1}\,p/\rho$, where $k$ is the Boltzmann constant and
$\bar{\mu}=1.28\rm\,amu$ the mean molecular mass.

\subsection{Chemistry and dust formation}
%========================================
\label{sec:dust}

The time-dependent formation of dust in the carbon-rich case is
calculated according to the moment method developed by Gail\plus
Sedlmayr (1988), Gauger\etal (1990) and Dominik\etal (1993). Assuming
spherical dust particles, the moments of the dust size distribution
function $f(V,\rr,t)$ are defined as \nocite{gs88,ggs90,dsg93}
\begin{equation}
  \rho L_j(\rr,t) = \int_{\Vl}^{\infty}\!\! f(V,\rr,t)\,V^{j/3}dV
                      \quad\mbox{$\big(j\in\{0,1,2,3\}\big)$}\ ,
\end{equation} 
where $V$ is the dust particle volume and $\Vl$ the minimum volume of
a large molecule to be counted as dust particle. The temporal change
of these dust moments, as affected by nucleation, growth and
evaporation, are calculated according to Eq.\,(4). Here, $J_\star\,[\rm
cm^{-3}s^{-1}]$ is the nucleation rate (\ie the creation rate of seed
particles) which is calculated for pure carbon chain molecules according to
Gail\etal (1984)\nocite{gks84}. $\chi^{\rm net}\,[\rm cm/s]$ is the
net growth velocity of already existing dust surfaces by the
accretion and thermal evaporation of carbon and hydrocarbon
molecules. The functional dependencies of these quantities are
\begin{eqnarray}
  \hspace*{25mm}
  J_\star &=& J_\star(\Tg,n_k)\\
  \chi^{\rm net} &=& \chi^{\rm net}(\Tg, \Td, n_k)\\
  n_k &=& n_k(\rho,\Tg,\epsilon_{\rm C})
\end{eqnarray} 
The various particle densities $n_k$ entering into this description
are calculated by assuming chemical equilibrium in the gas phase among
H, H$_2$, C, O, CO, H$_2$O, C$_2$, C$_3$, CH, C$_2$H, C$_2$H$_2$,
CH$_4$ according to the local mass density $\rho$, the local gas
temperature $\Tg$, and the local carbon element abundance
$\epsilon_{\rm C}$ in the gas phase. The latter obeys the following
conservation law
\begin{equation}
  \epsilon_{\rm C} \,+\, \frac{1.427\,{\rm amu}}{V_0} L_3 
  \,=\, \epsilon_{\rm C}^0 \,=\,\rm const
\end{equation}
$V_0\!=\!8.78\cdot 10^{-24}\rm cm^3$ is the monomer volume of
graphite and $1.427\,{\rm amu}$ the conversion factor between mass
density $\rho$ and hydrogen nuclei density $\nH$ for solar abundances. 
All other element abundances than carbon are assumed to have solar values.
We take the oxygen abundance $\epsilon_{\rm O}\!=\!10^{8.87-12}$ from 
(Grevesse\plus Noels 1993)\nocite{gn93} and consider the carbon-to-oxygen
ratio ${\rm C/O}\!=\!\epsilon_{\rm C}^0/\epsilon_{\rm O}$ as a free
parameter.

Dust evaporation is only included in terms of negative $\chi^{\rm
net}$ in case of under-saturation. Negative particle fluxes in size
space through the lower integration boundary at $\Vl$ are
neglected. For more details, see Gail\plus Sedlmayr (1988),
Gauger\etal (1990) and Dominik\etal (1993)\nocite{gs88,ggs90,dsg93}

\subsection{Radiative transfer}
%==============================
\label{sec:radtrans}

Radiative transfer in dusty winds is a complicated problem. The optical
depths are typically of the order of unity and the opaque regions can
be confined into shells or clouds which move in an otherwise
transparent medium. The opaque dust configurations scatter, re-emit
and cast shadows, thereby illuminating the optically thin regions in between
in complicated ways. Finding a efficient radiative transfer method to
resolve these problems is the key to develop multi-dimensional
models of dust-driven winds.

The absorption and thermal re-emission of radiation by dust grains
leads to a very fast relaxation of the internal dust temperature to
the local radiation field with cooling timescales of the order of
milliseconds for small amorphous carbon grains (Woitke\etal
1999)\nocite{woi99}. This fast relaxation introduces a stiff coupling
between hydrodynamics and radiative transfer. Any explicit scheme that
is based on formal solutions of the radiative transfer problem (\eg
short characteristics method) has the disadvantage to slow down the
computational timestep (typically a few $10^4$ seconds) to this
radiative cooling timescale, which is not acceptable.

We therefore seek for a numerical method that is capable to solve the
continuum radiative transfer problem under the auxiliary condition
that the dust component is in radiative equilibrium
\begin{equation}
  \int\! \kappa^{\rm dust}_{\rm abs} J_{\nu}\,d\nu =
  \int\! \kappa^{\rm dust}_{\rm abs} B_{\nu}(\Td)\,d\nu \ ,
\end{equation}
which de-stiffens the physical problem by the elimination of the
shortest characteristic timescale via assuming a quasi-equilibrium.
$\kappa^{\rm dust}_{\rm abs}$ is the dust absorption coefficient,
$J_{\nu}(\rr)=\frac{1}{4\pi}\!\int\!I_\nu(\rr,\nn)\,d^2n$ the mean
spectral intensity, $I_\nu(\rr,\nn)$ the spectral intensity in
direction of the unit vector $\nn$ and $B_{\nu}$ the Planck
function. In 2D or 3D, state-of-the-art Monte Carlo methods can do
this job with the same accuracy as ray-based methods at a similar
level of computational costs (\eg Pascucci\etal 2004).\nocite{pas2004}

We have therefore developed a new software module for the {\sc
Flash}-code which solves the LTE continuum radiative transfer with
radiative equilibrium in spherical coordinates by a Monte Carlo (MC)
method.  According to our knowledge, this is actually the first report
of such an approach in multi-dimensional fluid dynamics.

We have implemented the MC method of Niccolini, Woitke\plus Lopez
(2003)\nocite{nwl2003} which solves the fre\-quency-dependent
radiative transfer problem with angle-depen\-dent scattering. However,
for this paper, we will only consider the grey case with isotropic
scattering and simple fre\-quency-averaged opacities, because we want
to validate our new implementation by comparing the results of 1D
spherically symmetric winds to former publications, where these
opacities have been used (Fleischer\etal 1995, H{\"o}fner\etal 1995):
\begin{eqnarray}
  \kext &=& \widehat{\kappa}^{\rm\,ext}_{\rm gas} 
          + \widehat{\kappa}^{\rm\,ext}_{\rm dust}\\
  \widehat{\kappa}^{\rm\,ext}_{\rm gas}  &=& 2\cdot 10^{-4}\rm\,cm^2\,g^{-1}\\
  \widehat{\kappa}^{\rm\,ext}_{\rm dust} &=& \frac{3}{4} Q'(\Td)\,L_3
          \quad\mbox{with\ \ $Q'(\Td)=5.9\cdot\Td$}
\end{eqnarray}
The constant gas opacity is taken from Bowen (1988) and the
Rosseland mean opacity for small amorphous carbon grains from
Gail\plus Sedlmayr (1987).\nocite{bow88,gs87a}

Due to the extreme temperature-dependence of the nucleation and dust
evaporation process, the noise level in the temperature determination
should be as small as $\pm 1$\,K (see results in
Sect.~\ref{sec:stationary_winds}). The statistical error in our MC
method is determined by the number of crossing events of photon
packages through a cell under consideration. Therefore, in order to
achieve the required accuracy, one has to shoot many photon packages
{\it and} one cannot use too small computational cells.

This condition practically prevents a solution of the radiative
transfer problem on the partly very highly resolved AMR
hydro-grid. Instead, a comparably coarse regular grid (equally spaced
in $\log\,r$ and $\theta$) is used for the radiative transfer
calculations. A gas and dust mass conserving scheme has been
implemented, based on cell volume overlaps, to map the
$\{\rho,L_3\}(\rr)$ AMR hydro-data on this transfer grid. Next, the
opacities are calculated and the Monte Carlo experiment is carried
out, which results in the mean intensity structure
$J(\rr)=\frac{\sigma}{\pi}\Trad^4(\rr)$. This data is finally
interpolated back onto the hydro grid, using a 2D piecewise linear
interpolation scheme. The whole procedure is treated as additional
source term in the frame of the operator splitting method of the hydro
solver.

In order to achieve the required temperature accuracy of $\pm 1$\,K,
the computational cost for one radiative transfer call is quite
remarkable: about 40\,sec on one processor for a 200\,--\,grid (1D) and
about 120\,sec on 64 processors for a ($120\times 80$)\,--\,grid (2D). We
are therefore forced to call the MC radiative transfer routine 
less frequent than at every hydro timestep.

Instead, we have developed an extrapolation scheme that allows us to
call the radiative transfer routine only at about every 3$\rm^{rd}$
to 100$\rm^{th}$ hydro timestep. The basic idea for this extrapolation
in time is to interpolate in optical depth space. A parallel algorithm
to compute the spherical optical depths
\begin{equation}
  \tauL(r,\theta,t) = \!\!\int\limits_r^{R_{\rm out}}\!\!
                    \kappa_{\rm ext}(r',\theta,t)\, 
                    \bigg(\frac{\Rst0}{r'}\bigg)^2 dr'
\end{equation}
on the basis of the AMR data has been implemented, similar to
Rijkhorst\etal(2005)\nocite{rpdb2005}. $R_{\rm out}$ is the outer
boundary of the model domain and $\Rst0$ the stellar radius of the
initial model (see Sect.~\ref{sec:static}). Given the last radiative
transfer results $\Tref(\rr)=\Trad(\rr,t_{\rm ref})$ at reference time
$t_{\rm ref}$ with reference optical depths
$\tauref(\rr)=\tauL(\rr,t_{\rm ref})$, the actual radiation
temperatures are calculated as
\begin{eqnarray}
  \hspace*{1cm}
  \Trad^4(\rr,t) &=& \Tref^4(\rr)
     + a_{\rm cor}(\rr)\big(t-t_{\rm ref}\big) \nonumber\\
     &+& b_{\rm cor}(\rr)\Big(\tauL(\rr,t)-\tauref(\rr)\Big) \ ,
  \label{eq:Textrap}
\end{eqnarray}
where $a_{\rm cor}$ and $b_{\rm cor}$ are local fit coefficients which
are updated after every radiative transfer call.  $b_{\rm
cor}(\rr)=\pabl{\Trad^4}{\tauL}\Big{\vert}_{\theta, t_{\rm
ref}}\!\!\!\!\!(r)$ is the local dependence of $\Trad^4$ from
$\tauL$, which expresses typical backwarming effects. These effects
can occur as fast as hydrodynamical changes, for instance if an opaque
dust cloud enters a computational cell.  $b_{\rm cor}$ is calculated
by linear regression of the local $\{\Trad^4,\tauL\}$ data points
along the radial direction.

$a_{\rm cor}$ expresses the net change caused by all other radiative
transfer effects, e.g. by the passing of a dust cloud sideways of a
considered cell or by the changing shadow of a growing dust cloud between
the star and the cell. These changes are usually caused by the more
distant opacity configurations and hence occur on much longer
timescales. After having fixed $b_{\rm cor}$, the second fit
coefficient $a_{\rm cor}$ is calculated from inversion of
Eq.\,(\ref{eq:Textrap}), such that the extrapolation from the 
old to the new radiative transfer results would have been perfect.

The time interval $\Delta t_{\rm MC}$, after which the next proper MC
radiative transfer must be calculated and the fit coefficients $a_{\rm
cor}$ and $b_{\rm cor}$ are renewed, is controlled by the condition
that the difference between the extrapolated $\Trad$-values and the
newly calculated $\Tref$ MC results (measured by the 4-norm) must be
smaller than $2\times \Delta_{\rm MC} T$, where $\Delta_{\rm MC}
T\approx 1\rm\,K$ is the Monte Carlo noise in the temperature
determination.

We note that Eq.~(\ref{eq:Textrap}) allows for an exact treatment
of the following two limiting cases
\begin{equation}
  \begin{array}{lccc}
  \Tref^4 = \displaystyle \frac{1}{2}\,T_{\rm eff}^4 , &
  a_{\rm cor} = 0 , &
  b_{\rm cor} = \displaystyle \frac{3}{4}\,T_{\rm eff}^4 , &
  \tauref     = 0 \\
  \multicolumn{4}{c}{\Rightarrow \;\mbox{spherical diffusion approximation}}\\
  \multicolumn{4}{c}{\mbox{(Eddington approximation)}}\\[2mm]
  \Tref^4     = W(r)\,T_\star^4 , &
  a_{\rm cor} = 0 , &
  b_{\rm cor} = \displaystyle \frac{3}{4}\,T_\star^4\,
                \bigg(\frac{R_\star}{\Rst0}\bigg)^2 , &
  \tauref     = 0 \\
  \multicolumn{4}{c}{\Rightarrow \;\mbox{spherical two-stream approximation}}\\
  \multicolumn{4}{c}{\mbox{(Lucy's approximation\,,)}}
  \end{array}\nonumber
\end{equation}
where
$W(r)=\frac{1}{2}\big[1-\big(1\!-\!\frac{R_\star^2}{r^2}\big)^{1/2}\big]$
is the radial dilution factor.  By fixing $\Tref$, $a_{\rm cor}$ and
$b_{\rm cor}$ in this way, or by keeping them updated according to the
actual values of the stellar radius $R_\star(t)$ and stellar
temperature $T_\star(t)$, respectively, one can run quick test models
without involving the expensive MC radiative transfer routine. These
simplifications, however, are only possible for spherically symmetric 1D
models. 2D models require the proper inclusion of the MC radiative
transfer.

Equation (\ref{eq:Textrap}) is hence a reasonable approximation with
the ability to self-adapt to certain limiting cases if the MC results
indicate so. In general, it will only be valid in a local sense
for a limited time span after a last proper solution of the radiative
transfer problem. The extrapolated radiative transfer results are
always time-resolved because they depend on the actual optical depths.

Regarding the MC radiative transfer routine itself, scattering and
thermal re-emission are numerically treated in a different way
(Niccolini, Woitke\plus Lopez 2003)\nocite{nwl2003}. However,
re-emission and isotropic scattering is physically indistinguishable
in the grey case, so we can treat the albedo $\gamma$ as free
numerical parameter. Best performance is achieved by choosing
$\gamma=0.9$.

Another troublesome point is the $\Td$-dependence of
$\widehat{\kappa}^{\rm\,ext}_{\rm dust}$ which makes necessary an
iterative procedure between opacity and radiative transfer
calculations.  We solve this problem by calling the MC routine only
once (if it is foreseen for this timestep) and then using
Eq.\,(\ref{eq:Textrap}) in the subsequent iterations. After typically
1-5 iterations, this procedure converges (opacity calculation --
optical depths calculation -- $\Trad$ recalculation). Only if it does
not converge (which happens very seldom), more MC radiative transfer
calls within this iteration loop are required.

\subsection{Boundary conditions}
%===============================

Finding suitable boundary conditions for the multi-dimensional stellar
wind problem in Eulerian coordinates (fixed volume) is not trivial. The
main difficulty arises from the mass loss through the outer boundary
which, on average, should be compensated for by a mass inflow at the inner
boundary. The {\sc Flash}-code uses 4 layers of guard cells to
describe boundary conditions. On these guard cells, all variables
($\rho$, $\vv$, $e$, $L_j$) must be prescribed at every computational
time step.

\subsubsection{Inner boundary condition}
%=======================================

The basic idea of our permeable inner boundary is to fix the gas
pressure at the inner boundary. If matter is removed from the regions
close to the inner boundary, \eg blown away by radiation pressure, a
pressure gradient across the inner boundary develops which causes the
desired mass inflow.

In order to quantify this idea, we visit Eqs.\, (\ref{eq1}) and
(\ref{eq2}) in the spherically symmetric stationary case, where they
can be combined into the so-called wind equation
\begin{equation}
  v_r\pabl{v_r}{r} \left(1-\frac{c_T^2}{v_r^2}\right) 
  - \frac{2 c_T^2}{r} + \pabl{c_T^2}{r}
  = a_{\rm rad} - a_{\rm grav} \ .
  \label{eq:wind}
\end{equation}
$c_T\!=\!\sqrt{p/\rho}$ is the isothermal sound speed. Assuming\linebreak
$v_r\ll c_T$ (subsonic limit), Eq.\,(\ref{eq:wind}) becomes
\begin{equation}
  -\frac{1}{v_r}\pabl{v_r}{r} - \frac{2}{r} + \frac{1}{c_T^2}\pabl{c_T^2}{r}
  = \frac{a_{\rm rad} - a_{\rm grav}}{c_T^2} \ .
  \label{eq:wind2}
\end{equation}
By means of $\rho\,v_r r^2\!=\!\rm const$ (the stationary case of
Eq.\,\ref{eq1}) one can show that the l.h.s. of Eq.\,(\ref{eq:wind2})
equals $\pabl{}{r} \ln\,(\rho\,c_T^2)$, resulting in the integral
\begin{equation}
  \ln\,(\rho\,c_T^2)\Big\vert_r = \ln\,(\rho\,c_T^2)\Big\vert_{r_0}
  + \int\limits_r^{r_0} \frac{a_{\rm rad} - a_{\rm grav}}{c_T^2}\,dr \ ,
  \label{eq:inner_bc}
\end{equation}
which shows that stationary flows in the subsonic limit in fact obey the
hydrostatic condition.  We use Eq.\,(\ref{eq:inner_bc}) to describe
our inner boundary condition. We numerically evaluate the integral in
(Eq.\,\ref{eq:inner_bc}) from an arbitrary point $r$ on the guard
cells to the first valid cell within the model volume $r_0$ by
assuming that the gas temperature $\Tg$ can be linearly extrapolated
backward from the first two valid cells in the radial direction.

One way to fix the gas pressure at the inner boundary would be
$(\rho\,c_T^2)\big\vert_{r_0}\!\!=p(r_0,t\!=\!0)$. However, this
simple idea leads to the problem that a sudden increase of the
temperature level close to the inner boundary (\eg due to a dust shell
formation event) leads to a likewise sudden increase of the gas
pressure level, which can result in a {\it positive} pressure gradient
across the inner boundary, causing a vivid {\it infall} (negative
velocities) through the inner boundary. According to our experience,
it is better to fix the gas density at the inner boundary by
\begin{equation}
  (\rho\,c_T^2)\Big\vert_{r_0} = \rho(r_0,t\!=\!0)\;c_T^2(r_0,t) \ ,
\end{equation}
which leads to a more regular increase of the pressure level around
the inner boundary (also on the guard cells) in case of a sudden irradiation 
from the outside. A positive temperature offset lowers $|\pabl{\ln
p}{r}|$, \ie it causes a modest {\it outward acceleration} of the gas
in case of a sudden irradiation from the outside, which seems more
natural regarding a star to be characterised by an equilibrium between
gravity and pressure forces.

Once having solved Eq.\,(\ref{eq:inner_bc}) for $\rho$, the radial
velocity on the guard cells $v_r$ is calculated from $\rho\,v_r
r^2\!=\!\rm const$, the internal energy $e$ is calculated from $\rho$
and $\Tg$ via the equation of state, and the guard cells at the inner
boundary are assumed to be dust-free $L_j(r\!<\!r_0)\!=\!0$.

Since we do not consider stellar pulsation in this paper, a constant
stellar luminosity $L_\star$ is assumed as inner boundary condition
of the radiative transfer problem.

\subsubsection{Outer boundary condition}
%=======================================

For the outer boundary condition, a standard outflow condition
(zero-gradient, \ie a direct copy from the last valid cell designated by
the subscript $0$) is applied to the variables $\vv$, $\Tg$ and
$L_j$. The gas pressure is extrapolated in such a way that the
gradient $\pabl{p}{r}$ compensates gravity and radiative acceleration
(force-free):
\begin{equation}
  \ln\,p\Big\vert_r\! = \ln\,p_0
  + \frac{1}{c_{T_0}^2}\bigg(G M_{r_0}\!- \frac{\kext L_\star}{4\pi\,c}\bigg)
   \bigg(\frac{1}{r_0}-\frac{1}{r}\bigg) \ .
  \label{eq:outer_bc}
\end{equation}
Equation~(\ref{eq:outer_bc}) can be deduced from
Eq.\,(\ref{eq:inner_bc}) under the assumption that $M_r$, $c_T$ and
$\kext$ are constant on the guard cells. Mass density $\rho$ and
internal energy $e$ are finally calculated from $p$ and $\Tg$, using
the equation of state.

\subsubsection{Angular boundary conditions}
%==========================================
In case of a 2D model, the guard cells at $\theta\!<\!0$ and
$\theta\!>\!\pi$ are filled by a standard reflecting boundary
condition which assures $v_\theta(r,\theta\!=\!0,t) =
v_\theta(r,\theta\!=\!\pi,t)=0$.

\subsection{Initial condition}
%=============================
\label{sec:static}
The dynamical simulations are started from a dust-free static stellar
atmosphere as initial condition. In the static (and hence spherically
symmetric) case, the Poisson equation, the equation of motion and the
definition of the optical depth form a system of ordinary differential
equations
\begin{eqnarray}
  \abl{M_r}{r}   &=& 4\pi r^2 \rho 
                 \label{eq:static1}\\
  \abl{p}{r}     &=& \rho\,\bigg(\frac{\kext L_\star}{4\pi r^2 c} 
                               - \frac{G M_r}{r^2}\bigg) 
                 \label{eq:static2}\\
  \abl{\tauL}{r} &=& -\rho\,\kext\bigg(\frac{\Rst0}{r}\bigg)^2
                 \label{eq:static3}
\end{eqnarray}
which is completed by the equation of state $\rho = \rho(p,\Tg)$, the
gas opacity $\kext = \widehat{\kappa}^{\rm\,ext}_{\rm gas}(\rho,\Tg)$
and an analytic formula of the temperature as function of optical
depth $\Tg = \Tg(r,\tauL)$. This formula represents the solution of
the radiative transfer problem in radiative equilibrium.

Starting with an initial guess of this dependency,
Eqs.~(\ref{eq:static1}) to (\ref{eq:static3}) can be solved by a
standard ODE solver with the outer boundary condition
\begin{eqnarray}
  \tauL(R_{\rm out}) = 0
\end{eqnarray}
and two conditions at the stellar radius 
\begin{eqnarray}
  M_r(\Rst0) &=& M_\star \\
  \Tg(\Rst0) &=& T_{\rm eff} \ ,
\end{eqnarray}
where the initial stellar radius $\Rst0$ is given by the
Stefan-Boltzmann law $L_\star = 4\pi {\Rst0}^2 \sigma T_{\rm
eff}^{\,4}$.

Having obtained this solution, the MC radiative transfer routine is
called for given gas density distribution $\rho(r)$ which provides new
$\{r,\Tg,\tauL\}$ data points, from which an updated version of the
temperature formula (piecewise linear interpolation) is
calculated. The procedure needs about 3-10 iterations to converge.

The physical parameters of the complete static stellar atmosphere
problem are hence the stellar luminosity $L_\star$, the effective
temperature $T_{\rm eff}$ and the stellar mass $M_\star$.  The static
stratification is calculated during the initialisation phase of the
{\sc Flash}-hydrocode. The solution is then mapped onto the initial
{\sc Flash}-grid, again using a piecewise linear interpolation scheme
in $r$.

\subsection{Timestep control}
%============================
\label{sec:timestep}
The computational timestep of the dust radiation hydrodynamics code is
limited by two criteria. First, $0.8\times$\,the Courant-Friedrich-Levy
timestep, given by the minimum of the sound wave travel time through any 
computational cell, and second, a temperature limiter. The latter
limits the maximum change of $\Trad$ during one computational
timestep to some given value. We choose this value to be $5\,$K, which
seems necessary for the simulation of dust formation and in fact
stabilises the results.

\subsection{Model domain and grid resolution}
%============================================
\label{sec:grid}
As radial model domain, the interval $r/\Rst0\in[0.9,10]$ is considered.
In case of a 2D-model, the full angular range $\theta\in[0,\pi]$ is
considered. The inner boundary is chosen to be located at
$\tauL\approx 5$, where the velocity variations are truly subsonic and
the radiative transfer is diffusive. The outer boundary is located far
outside the dust formation and wind acceleration zone. We have checked
that a larger choice than $10\,\Rst0$ has only minor influence on the
resulting wind quantities like \eg the mass loss rate.

We start with a basic grid resolution (refinement level~1) of
$[r\!\times\!\theta] = 96\times 64$ cells. Based on second derivative
criteria, the grid will be refined during the simulation to maximum
level 5. Each refinement step increases the local resolution
block-wise by a factor of two in each spatial dimension, such that the
maximum resolution (if the grid needs to be fully refined) is
$1536\times 1024$.

\subsection{Computational limitations}
%=====================================
\label{sec:limits}
The implementation of the physical and chemical processes (source
terms) in the frame of the {\sc Flash}-solver is independent
of spatial dimension and applied coordinate system, such that now
1D/2D/3D {\sc Flash} simulations with Cartesian or curvilinear
coordinates are principally possible. We find it, however, most
convenient, if not even mandatory, to formulate the boundary
conditions with spherical coordinates. Furthermore, the Monte Carlo
radiative transfer code does currently only allow for 1D and 2D models
with spherical coordinates.

A complete high-resolution 1D model (60 years of simulation) requires
about $5\times 10^5$ CPU sec, a complete 2D high-resolution model
about $5\times 10^7$ CPU sec on {\sc Aster}, which is an SGI Altix
3700 system consisting of 416 processors (Intel Itanium~2, 1.3
GHz). We have typically used 8 processors for a 1D model (16 hours
user time) and 64 processors for a 2D model (9 days). Hence, even with
the present power of modern parallel super-computers, 3D simulations
with Monte Carlo radiative transfer are simply too expensive for the
required accuracy (see Sect.~\ref{sec:stationary_winds}).  Thus, for
practical purposes, only 1D (spherically symmetric) and 2D
(axisymmetric) simulations can be run with the present code.

%----------------------------------------------------------------------------
\section{Results}
%----------------------------------------------------------------------------

The description of our results is separated into two parts. First, we
compare our results for spherically symmetric (1D) dust-driven winds
with other published works in Sect.~\ref{sec:results_1D}, intending to
validate our new numerical implementation. In
Sect.~\ref{sec:results_2D}, the new axisymmetric (2D) models will be
presented and discussed.

%=======================================
\subsection{1D results: code validation}
%=======================================
\label{sec:results_1D}

\begin{table}
\centering
\caption{Results of 1D spherically symmetric models for C/O increase
(see text). Constant parameters: $M_\star=1\,M_\odot$ and
$L_\star=10^4L_\odot$.}
\label{tab1}
\vspace*{-1mm}
\begin{tabular}{cl|ccc|c}
$\!\!\!\!T_{\rm eff}\rm[K]\!\!\!\!$ 
                        & \!\!C/O & $\dot{M}[\frac{M_\odot}{\rm yr}]$\!\! 
                        & \!\!\!\!\!$v_\infty[\rm km/s]$\!\!\!\!\!\!\!
                        & \!\!$\rho_{\rm d}/\rho_{\rm g}$\!\!
                        & remarks\\
&&&&&\\[-2.5ex]
\hline
  2400 & 1.6   & $\approx 3(-7)$ &  $<2$  & $\approx 8(-4)$ & breeze \\
  2400 & 1.65  &       $4.6(-7)$ &   4.1  & $8.1(-4)$  & stationary \\
  2400 & 1.7   &       $6.3(-7)$ &   6.1  & $8.6(-4)$  & stationary \\
  2400 & 1.75  &       $1.0(-5)$ &  24.6  & $2.7(-3)$  & $P=790\,$d \\
  2400 & 1.8   &       $1.1(-5)$ &  25.2  & $2.7(-3)$  & $P=790\,$d \\
  2400 & 1.9   &       $9.9(-6)$ &  26.8  & $2.9(-3)$  & irregular  \\
\hline
  2500 & 1.8   & $\approx 1(-7)$ &  $<2$  & $\approx 1(-3)$ & breeze \\
  2500 & 1.85  &       $2.3(-7)$ &   5.1  & $8.1(-4)$  & stationary\\ 
  2500 & 1.9   &       $3.1(-7)$ &   6.8  & $8.6(-4)$  & stationary\\
  2500 & 1.95  &       $4.0(-7)$ &   8.7  & $9.5(-4)$  & stationary\\
  2500 & 2.0   &       $5.5(-7)$ &  11.5  & $1.3(-3)$  & stationary$^{(1)}$\\
  2500 & 2.05  &       $6.9(-6)$ &  27.5  & $2.7(-3)$  & $P=660\,$d \\
  2500 & 2.1   &       $6.2(-6)$ &  30.7  & $3.1(-3)$  & irregular\\
  2500 & 2.2   &       $7.9(-6)$ &  31.7  & $3.2(-3)$  & irregular\\
\hline
  2600 & 2.0   & $\approx 3(-8)$ &  $<2$  & $\approx 9(-4)$ & breeze \\
  2600 & 2.1   &       $7.1(-8)$ &   4.2  & $7.8(-4)$  & stationary\\
  2600 & 2.2   &       $1.2(-7)$ &   7.2  & $8.6(-4)$  & stationary\\
  2600 & 2.3   &       $1.9(-7)$ &  10.2  & $9.8(-4)$  & stationary\\
  2600 & 2.4   &       $3.8(-6)$ &  32.0  & $3.3(-3)$  & irregular\\
  2600 & 2.5   &       $3.9(-6)$ &  35.5  & $3.4(-3)$  & irregular\\
\end{tabular}\\[2mm]
{\footnotesize \begin{tabular}{p{10mm}p{71mm}}
breeze:  & Subsonic solution with finite outer pressure for $r\!\to\!\infty$
           (unphysical). In the numerical model, such solutions
           are featured by a standing shock wave in the outer regions.\\ 
\hbox{$P=...$} & Quasi-periodic solution. All global quantities like 
           $\dot{M}(t)$ or $M_{\!d}(t)$ show cyclic variations (see 
           Fig.~\ref{fig:1Dtypes}). The Fourier 
           spectra are featured by an outstanding peak at a certain period 
           $P\,$[days]. Dust is produced at every cycle, but small 
           cycle-to-cycle variations may exist.\\
irregular: & Oscillatory solution with irregular dust shell production. 
           No clear period. Sometimes period switches.\\ 
$^{(1)}$:  & Stationary solution with small oscillatory perturbations
           (very close to the bifurcation point).\\
\end{tabular}}
\vspace*{-3mm}
\end{table}

Fleischer\etal (1995) and H{\"o}fner\etal (1995)\nocite{fgs95,hfd95}
have published spherically symmetric (1D) models for time-dependent
dust-driven winds of carbon stars with the same physical equations and
assumptions (\eg no pulsation) and parameters (\eg opacities) as in
our model, but using different numerical techniques. These results can
hence be used to validate our new computational implementation. We
will briefly summarise the basic results obtained by all
three models, before we can discuss the differences.

After the start of a new simulation, the initially hydrostatic and
dust-free stratification of the stellar atmosphere and the
circumstellar environment undergoes some dramatic changes with strong
shock waves in the outer regions. However, the model structure soon
relaxes towards a new type of solution, which is found to be either a
{\it stationary wind} or an {\it oscillating wind}.

\begin{figure}
\centering
\epsfig{file=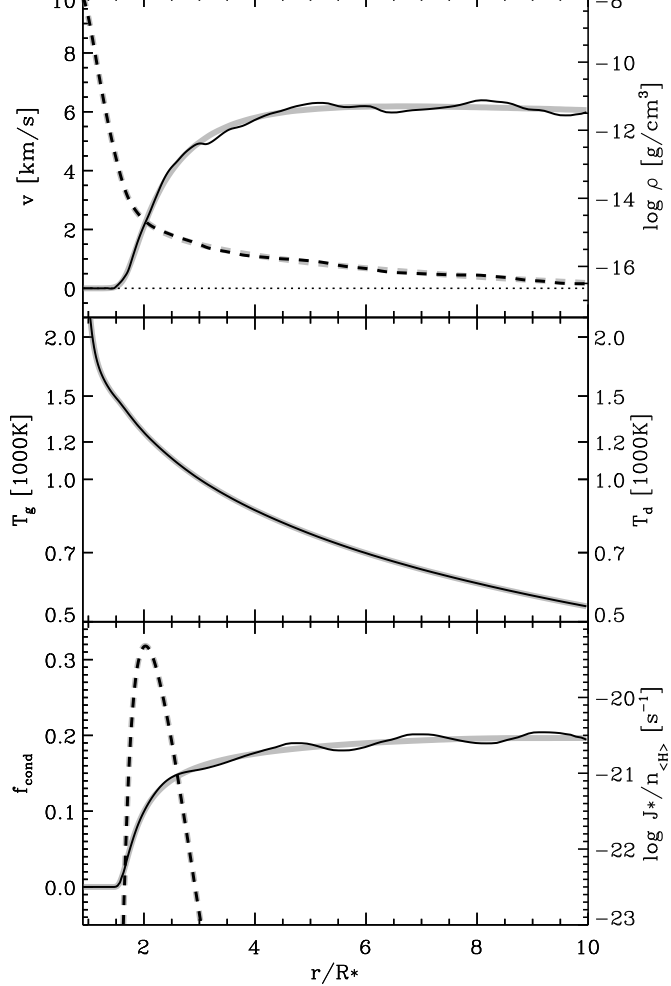, width=8.8cm}
\vspace*{-5mm}
\caption{Stationary 1D wind solution for $M_\star=1\,M_\odot$,
  $L_\star=10^4L_\odot$, $T_{\rm eff}=2400\,$K and $\rm C/O=1.7$. The
  thin black lines show the velocity $v$ (full), the mass density $\rho$
  (dashed), the dust and gas temperatures $\Td$ and $\Tg$ (full --
  practically identical in this model), the degree of condensation
  $f_{\rm cond}$ (full) and the nucleation rate per hydrogen nucleus
  $J_\star/\nH$ (dashed).  The underlying thick grey lines show the
  results for a comparison model, where the Monte Carlo radiative
  transfer code is not called anymore (see text).  The difference
  between black and grey indicates the effect of the Monte Carlo
  radiative transfer noise ($\Delta T_{\rm MC}\approx 1\,$K) on the
  solution.}
\label{fig:1D_stationaer}
\vspace*{-1mm}
\end{figure}

\begin{figure}
\centering
\epsfig{file=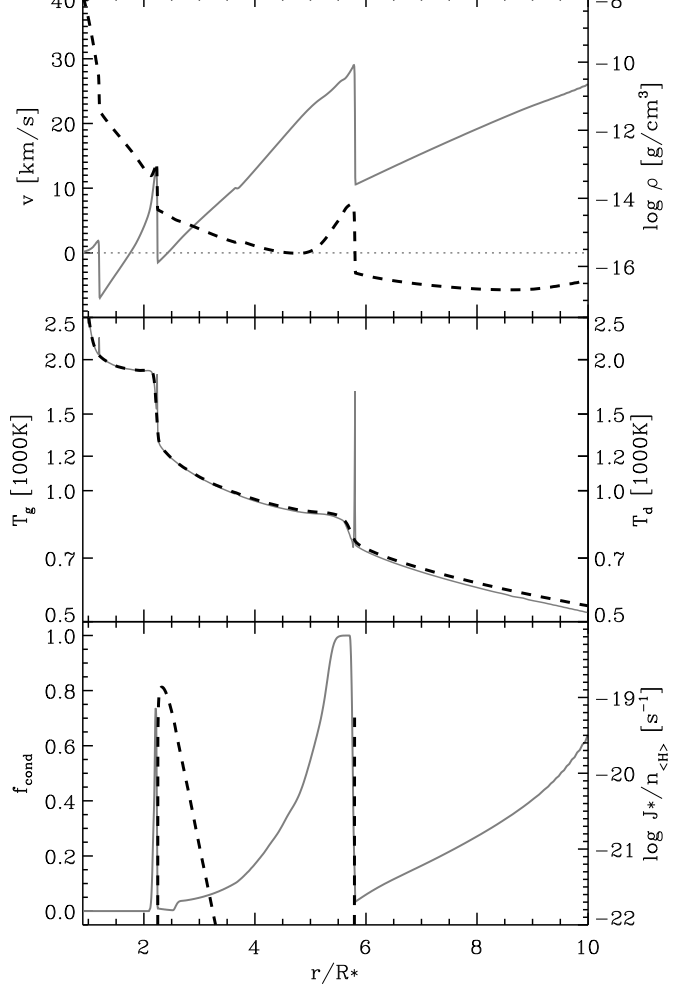, width=8.8cm}
\vspace*{-5mm}
\caption{Snap shot of an 1D oscillating wind solution for the same
  stellar parameters as in Fig.~\ref{fig:1D_stationaer} but with
  slightly increased $\rm C/O=1.75$.  In this plot, grey full lines
  refer to l.h.s. quantities and black dashed lines to r.h.s. quantities.}
\label{fig:1D_oscillatory}
\vspace*{-1mm}
\end{figure}

\subsubsection{Stationary winds}
%===============================
\label{sec:stationary_winds}
The stationary solutions are featured by low mass loss rates
$\dot{M}$, low outflow velocities $v_\infty$ and low dust-to-gas
ratios $\rho_{\rm d}/\rho_{\rm g}$ and occur for comparably small C/O
ratios (see Table~\ref{tab1}). Figure~\ref{fig:1D_stationaer} shows an
example for such a finally almost time-independent solution.  The inner
regions, from the bottom of the atmosphere to the start of the dust
formation zone, are approximately in hydrostatic equilibrium.  The
density falls off exponentially here and the velocity is approximately
zero. At some distance from the star (here at about $1.5\,R_\star$)
the temperature becomes low enough for nucleation which triggers the
process of dust formation. Radiation pressure on dust then accelerates
the dust/gas mixture and drives it through the sonic point located at
about $2\,R_\star$ in the depicted model. The wind region beyond that
point is featured by an almost constant outflow velocity and a
$\rho\propto r^{-2}$ law.  Due to the increasingly rapid outflow and
the decreasing gas densities, the process of dust formation freezes
in. For further details, see \eg Gail\plus Sedlmayr (1987).

Note that the formation of dust remains {\rm incomplete}. Only about
18\% of the condensable carbon has actually condensed into solid
particles in the depicted model (see $f_{\rm cond}$ in
Fig.~\ref{fig:1D_stationaer}). The influence of the dust formation on
the radiative transfer results is small. The model describes an
optically thin, low-velocity, dust-driven stellar wind.

\begin{figure*}
\hspace*{-3mm}
\begin{tabular}{cc}
\epsfig{file=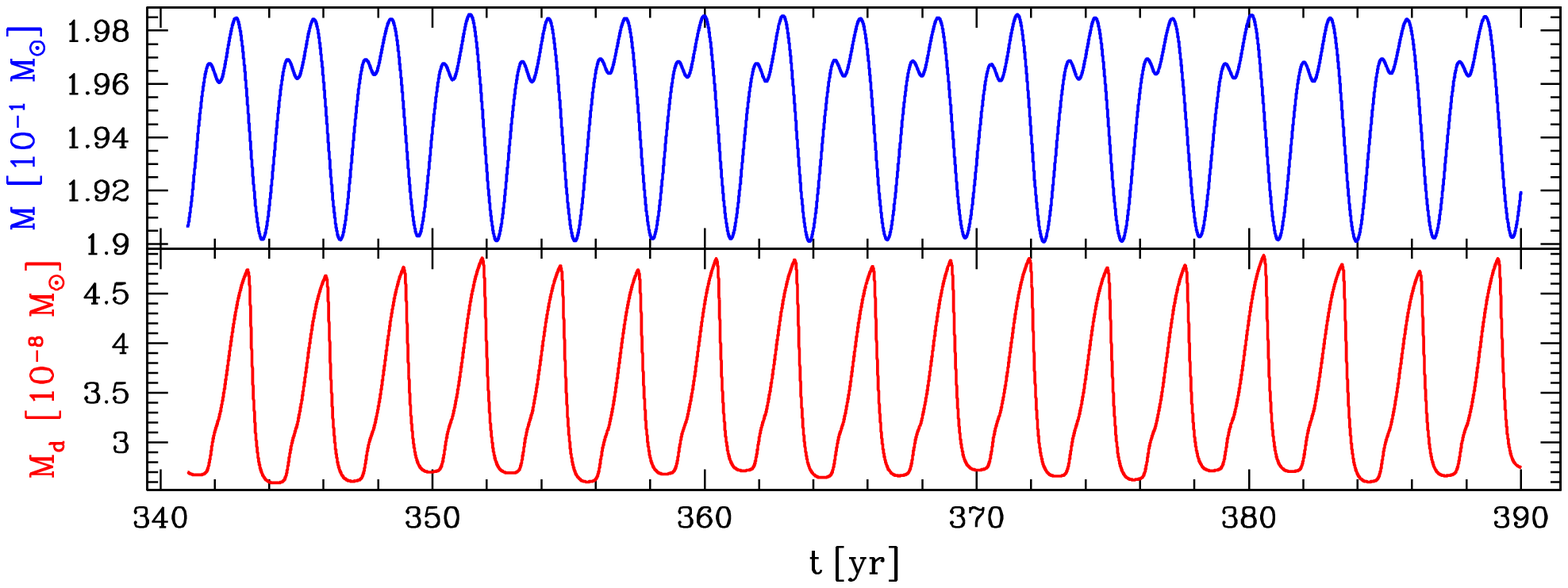, width=8.8cm}&
\epsfig{file=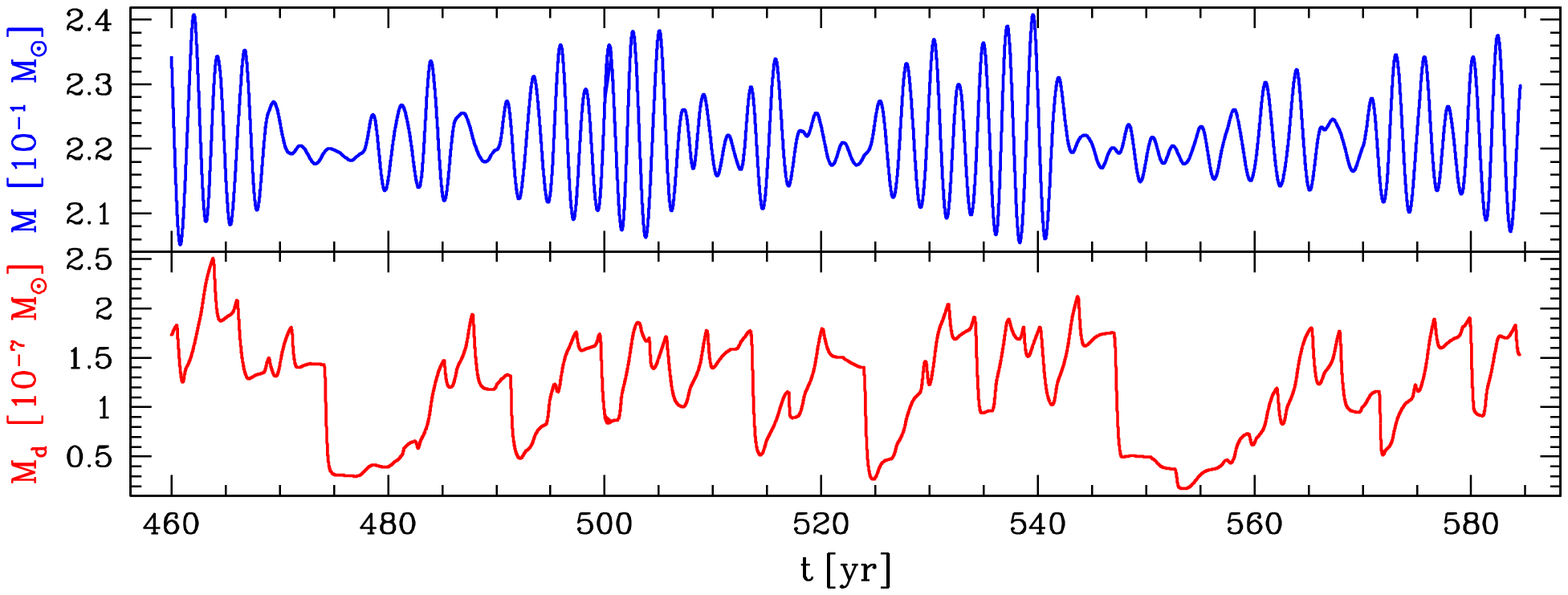, width=8.8cm}
\end{tabular}
\vspace*{-2mm}
\caption{Examples for ``quasi-periodic'' (l.h.s.: $T_{\rm
  eff}\!=2500\,$K and $\rm C/O\,{\scriptstyle\searrow}\,1.6$) and
  ``irregular'' (r.h.s.: $T_{\rm eff}\!=2400\,$K and $\rm
  C/O\,{\scriptstyle\searrow}\,1.6$) 1D oscillatory wind solutions.
  Depicted is the total $\rm gas\!+\!dust$ mass $M(t)$ (upper row) and
  the total dust mass $M_{\!\rm d}(t)$ (lower row) contained in the
  model volume. The change of $M(t)$ is mainly due to mass inflow and
  outflow through the inner boundary. The change of $M_{\!\rm d}(t)$
  is due to dust formation and mass loss through the outer
  boundary. In the quasi-periodic model (l.h.s.), a new shell of dust
  is formed at every second oscillation cycle (double-periodic
  solution). Fourier analysis shows two sharp maxima at 1060 and
  530 days. The basic period $P\!=\!530$ days given in Table~\ref{tab1}
  refers to the oscillation of the stellar atmosphere. The irregular
  model (r.h.s.) shows the production of dust shells at almost
  unpredictable instants of time.}
\label{fig:1Dtypes}
\vspace*{-1mm}
\end{figure*}

Due to the Monte Carlo noise in the temperature determination of our
radiative transfer code (see Sect.~\ref{sec:radtrans}), the relaxation
towards a stationary wind is actually never complete. The slightly
random radiative transfer results cause small temperature changes in
the inner regions which create inward and outward travelling sound
waves.  The outward travelling waves amplify in the large density
gradient close to the star. This amplification of waves and the strong
temperature dependence of the nucleation process result in noticeable
variations of the fluid variables in the outer regions of order 5\%,
although the temperature noise $\Delta T_{\rm MC}$ is as small as
0.1\% (not visible). This result leads us to the conclusion that an
accuracy of the radiative transfer results of $\Delta T_{\rm
MC}\approx 1\,$K is actually required to model dust-driven winds,
which makes the simulations very expensive.  We have checked that a
reduction of the Monte Carlo noise by a factor of two (by using four
times more photon packages) reduces the amplitudes of the remaining
disturbances by about a factor of two.

We can suppress the creation of new disturbances completely by not
calling the radiative transfer routine anymore, thereby fixing $\Tref$
and $b_{\rm cor}$ for ever and putting $a_{\rm cor}\!=\!0$ (see
Eq.~\ref{eq:Textrap})\footnote{This is not the same as fixing the
temperature profile: $\tau$-changes still cause $\Trad$-changes.}. In
that case, the stationary solution becomes completely smooth and
freezes in for all times (see underlying grey model in
Fig.~\ref{fig:1D_stationaer}).  This freezing for arbitrary long times
is possible, because our inner boundary condition compensates the mass
loss through the outer boundary, in contrast to the Fleischer\etal
(1995) and H{\"o}fner\etal (1995) models, where the model volume runs
empty for very long times.

\subsubsection{Oscillatory winds}
%================================
\label{sec:osc}
Beyond some critical value for the stellar luminosity (Fleischer\etal
1995) or the C/O ratio (H{\"o}fner et al. 1995), the stationary wind
solutions become unstable. This instability, called ``exterior
$\kappa$-mechanism'', is caused by feedbacks of the dust formation
process on the radiative transfer. After the formation of a new
optically thick dust shell, the temperatures inside of this shell
increase quickly due to {\it backwarming} (see $\Td(r)$ in
Fig.~\ref{fig:1D_oscillatory}), which switches off the nucleation
process and may even cause dust evaporation in the inner regions. The
newly formed dust shell is hence truncated at its inner edge and
driven outward, which reduces again its optical depth because of
radial dilution. Consequently, the temperatures in the inner regions
start to slowly decrease again, and a new cycle of dust formation may
begin.

This mechanism, however, interferes with other hydrodynamical
processes in the dust formation zone. The sudden acceleration of a
newly formed dust shell creates two waves. The first wave, an outward
travelling compression wave, steepens up into a shock wave,
compresses the gas ahead of the outward moving shell and
facilitates subsequent dust formation in the wake of this shock. The
second wave, an inward travelling expansion wave, causes the dust
formation zone to be re-filled with fresh matter from the star. The
sudden increase of the temperatures in the stellar atmosphere creates
a third wave, a pressure-driven expansion wave, which merges with the
incoming expansion wave to create a new shock wave below the dust
formation zone. These hydrodynamical disturbances affect the structure
of the stellar atmosphere and may interfere with the stellar pulsation
(disregarded in this paper). The characteristic timescale of the
dust-induced $\kappa$-mechanism (usually longer than typical
pulsational periods) may or may not coincide with the timescale of
these hydrodynamical processes. A complicated time-dependent behaviour
results in this way, which can be periodic, multi-periodic or completely
irregular (see Fig.~\ref{fig:1Dtypes}).

As a common result from all three simulations (this work,
Fleischer\etal 1995, H{\"o}fner\etal 1995), the oscillatory solutions
are featured by higher mass loss rates $\langle \dot{M}\rangle$,
higher outflow velocities $\langle v_\infty\rangle$ and higher
dust-to-gas ratios $\langle \rho_{\rm d}/\rho_{\rm g}\rangle$ as
compared to the stationary solutions (measured by temporal mean values
at the outer boundary). The solutions describe optically thick winds
of sometimes massively obscured central stars, \eg infrared carbon
stars.

The effect of the Monte Carlo noise on the oscillatory wind solutions
is more difficult to assess, because we cannot suppress it completely.
The ongoing creation of small temperature disturbances by the MC noise
in the stellar atmosphere prevents a complete relaxation to a truly
periodic solution. Therefore, our solutions are always featured by
small deviations from periodicity or are even irregular. This
behaviour, however, was also found by Fleischer\etal (1995) and
H{\"o}fner\etal (1995), who used deterministic radiative transfer
methods. If we use the deterministic Lucy approximation for the
radiative transfer (see Sect.~\ref{sec:radtrans}), we can produce
solutions for some parameter combinations which are sometimes closer
to truly periodic, but the general behaviour of the solutions is very
similar.  Therefore, we conclude that the Monte Carlo noise is not a
critical issue concerning the resulting type of wind solution.

%We have not been able to obtain a truly periodical
%solution as described in (H{\"o}fner et al. 1995), which is probably
%due to the ongoing creation of small temperature disturbances by the
%MC noise, which prevents a complete relaxation of the model.
%An increase of the number of photon packages (to reduce
%the MC noise) has usually little effect on the results. 
%If we use Lucy's approximation for the radiative transfer instead (see
%Sect.~\ref{sec:radtrans}), we can actually (for some parameter
%combinations) produce solutions which are closer to truly periodic.
%The overall model behaviour, however, is similar as reported by
%Fleischer\etal (1995) and H{\"o}fner\etal (1995), who used
%deterministic radiative transfer methods, and showed solutions with
%either small deviations from periodicity (``quasi-periodic'') or
%irregular behaviour. We conclude that the Monte Carlo noise is not a
%critical issue concerning the resulting type of wind solution.

\subsubsection{Comparison to other 1D models}
%============================================

The general behaviour of our new {\sc Flash} simulations for
dust-driven C-star winds in the special case of spherical symmetry
resembles well the behaviour reported by Fleischer\etal (1995) and
H{\"o}fner\etal (1995), progressing from breezes $\to$ stationary
winds $\to$ oscillatory solutions with increasing C/O. 
%Unfortunately,
%these authors have refrained from running models with identical
%parameters in these papers, making a detailed comparison difficult. 
In a comparative study, H{\"o}fner\etal (1996)\nocite{hfgfd1996} have
argued for good general agreement between the two codes despite the
different numerical techniques. Since Fleischer\etal (1995) have
considered throughout higher stellar luminosities as in this paper, we
will concentrate on the comparison to the H{\"o}fner et al.\ models in
the following.

\begin{figure}
\centering
\epsfig{file=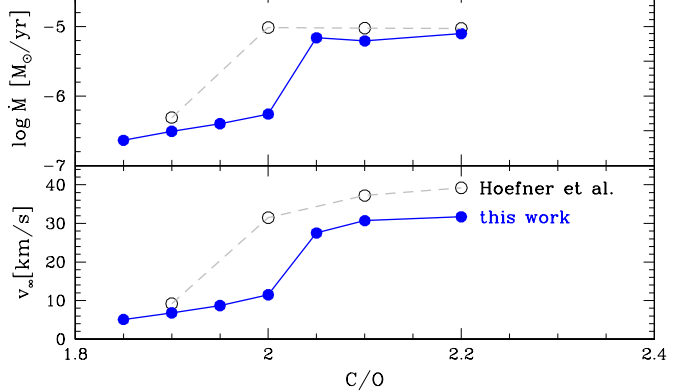, width=8.8cm}
\vspace*{-5mm}
\caption{Comparison of 1D results to H{\"o}fner\etal (1995) for 
  $M_\star=1\,M_\odot$, $L_\star=10^4L_\odot$ and $T_{\rm eff}\!=\!2500\,$K.}
\label{fig:vergl}
\vspace*{-1mm}
\end{figure}

Figure~\ref{fig:vergl} shows the differences of the calculated mean
wind quantities to H{\"o}fner\etal (1995, see their Table~3). In
summary, the H{\"o}fner et al.\ winds are a little stronger with slightly
larger $\dot{M}$ and $v_\infty$. One can also describe these
differences in the following way. The ``jump'' from stationary $\to$
oscillatory solutions in our models occurs at a slightly larger
critical C/O value as compared to the H{\"o}fner et al.\ models. The
obtained periods of H{\"o}fner et al.\ show the same trend (increasing
with increasing $T_{\rm eff}$) but are throughout shorter by a factor
of about 1.5.

The slightly lower efficiency of the driving by dust in our work can
already be explained by a tiny detail in the treatment of equilibrium
chemistry. In contrast to H{\"o}fner\etal (1995), we have included the
C$_3$ molecule which reaches high concentrations around $\approx
2000\,$K and thereby reduces the number density of C atoms and
C$_2$H$_2$ molecules. This lowers the efficiency of the nucleation
and growth of the carbon dust particles. According to our model, just
ignoring the C$_3$ molecule results in higher $\dot{M}$ and $v_\infty$
by a factor of about two in the stationary regime, which would put our
results even above those of H{\"o}fner et al.\ in Fig.~\ref{fig:vergl}.

However, various other numerical details could also be the reason for
the shown discrepancies. With regard to the very complex
physical system under investigation, small numerical deviations can
have big effects.  In view of the different radiative
transfer methods, the different implementation of the hydrodynamical
equations with different numerical schemes (implicit/explicit) and,
last but not least, the different inner boundary condition, the degree
of agreement is actually quite remarkable.

\subsubsection{The folding bifurcation}
%======================================

\begin{table}
\centering
\caption{Results of 1D spherically symmetric models for C/O decrease
(see text). Constant parameters: $M_\star=1\,M_\odot$ and
$L_\star=10^4L_\odot$.}
\label{tab2}
\vspace*{-1mm}
\begin{tabular}{cl|ccc|c}
$\!\!\!\!T_{\rm eff}\rm[K]\!\!\!\!$ 
                        & \!\!C/O & $\dot{M}[\frac{M_\odot}{\rm yr}]$\!\! 
                        & \!\!\!\!\!$v_\infty[\rm km/s]$\!\!\!\!\!\!\!
                        & \!\!$\rho_{\rm d}/\rho_{\rm g}$\!\!
                        & remarks\\
&&&&&\\[-2.5ex]
\hline
  2400 & 1.7   &       $1.1(-5)$ &  23.7  & $2.5(-3)$  & $P=790\,$d \\
  2400 & 1.6   &       $9.8(-6)$ &  19.0  & $1.8(-3)$  & irregular \\
  2400 & 1.5   &       $5.4(-6)$ &  18.0  & $1.6(-3)$  & irregular \\
  2400 & 1.4   &       $1.1(-6)$ &   3.4  & $7.8(-4)$  & osc.~breeze\\
                                                       % $P=900\,$d \\
  2400 & 1.3   &       $3.0(-7)$ &   0.4  & $8.9(-4)$  & osc.~breeze \\
\hline
  2500 & 2.0   &       $7.2(-6)$ &  30.0  & $3.0(-3)$  & irregular\\
  2500 & 1.9   &       $7.1(-6)$ &  25.2  & $2.3(-3)$  & irregular\\
  2500 & 1.8   &       $4.8(-6)$ &  21.0  & $1.9(-3)$  & irregular\\
  2500 & 1.7   &       $3.9(-6)$ &  21.2  & $1.6(-3)$  & irregular\\
  2500 & 1.6   &       $3.4(-6)$ &  20.7  & $1.6(-3)$  & $P=530\,$d$^{(2)}$\\
  2500 & 1.5   &       $5.8(-7)$ &   2.8  & $3.7(-4)$  & osc.~breeze\\
\hline
  2600 & 2.3   &       $4.6(-6)$ &  31.3  & $2.9(-3)$  & irregular\\
  2600 & 2.2   &       $4.2(-6)$ &  27.5  & $2.3(-3)$  & irregular\\
  2600 & 2.1   &       $2.8(-6)$ &  26.3  & $2.1(-3)$  & irregular\\
  2600 & 2.0   &       $1.2(-6)$ &  20.8  & $1.7(-3)$  & irregular\\
%  2600 & 1.9   &       $3.8(-7)$ &   9.5  & $1.1(-3)$  & irregular\\
  2600 & 1.8   &       $1.7(-6)$ &  20.9  & $1.7(-3)$  & $P=630\,$d$^{(2)}$\\
  2600 & 1.7   &       $2.5(-6)$ &  19.4  & $1.4(-3)$  & $P=620\,$d$^{(2)}$\\  
  2600 & 1.6   &       $1.6(-7)$ &   1.1  & $9.3(-4)$  & osc.~breeze \\
\end{tabular}\\[2mm]
{\footnotesize \begin{tabular}{p{14mm}p{67mm}}
osc.~breeze: & Low-velocity solution (sometimes subsonic $v_\infty$) with 
           small oscillations and persistent dust inside the nucleation regime 
           (see Fig.~\ref{fig:1D_breeze}).\\
$^{(2)}$:  & Double-periodic: formation of a dust shell at every second 
           oscillation cycle of the outer stellar atmosphere.        
\end{tabular}}
\end{table}

\begin{figure}
  \centering
  \epsfig{file=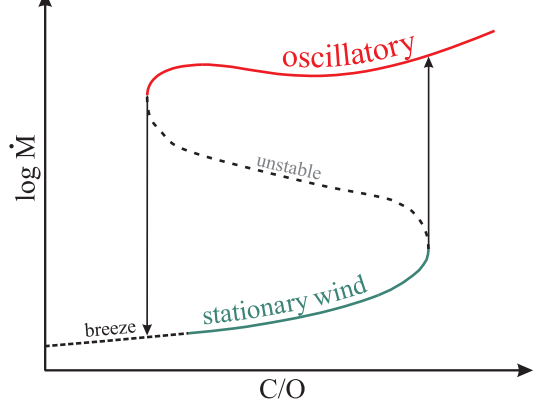, width=7cm}
  \vspace*{-2mm}
  \caption{Folding Bifurcation in modelling C-star winds.}
  \vspace*{-2mm}
  \label{fig:bifu}
\end{figure}

The results of our 1D simulations, as summarised in Table~\ref{tab1},
have been obtained by a stepwise increase of C/O, giving the model
sufficient time in between to relax to the respective equilibrated
solution (about 50\,yrs).

However, if we stepwise {\it decrease} C/O again, we face a surprise
(see Table~\ref{tab2}).  The model does not fall back toward the
stationary solution at the critical C/O value for increase as observed
before. Instead, it stays oscillatory until it finally falls back to a
``breeze'' at another, much lower critical C/O value for
decrease. Apparently, if once the transition to an oscillatory mode is
made and the outer parts of the stellar atmosphere are swinging back
and forth considerably, a more efficient mechanism of dust production
can be sustained even with less fuel. This non-uniqueness for
intermediate C/O values (hysteresis behaviour / folding bifurcation) is
sketched in Fig.~\ref{fig:bifu} and has not been reported so far.

\subsubsection{Oscillating low-velocity solutions}
%=================================================

\begin{figure}
  \centering
  \epsfig{file=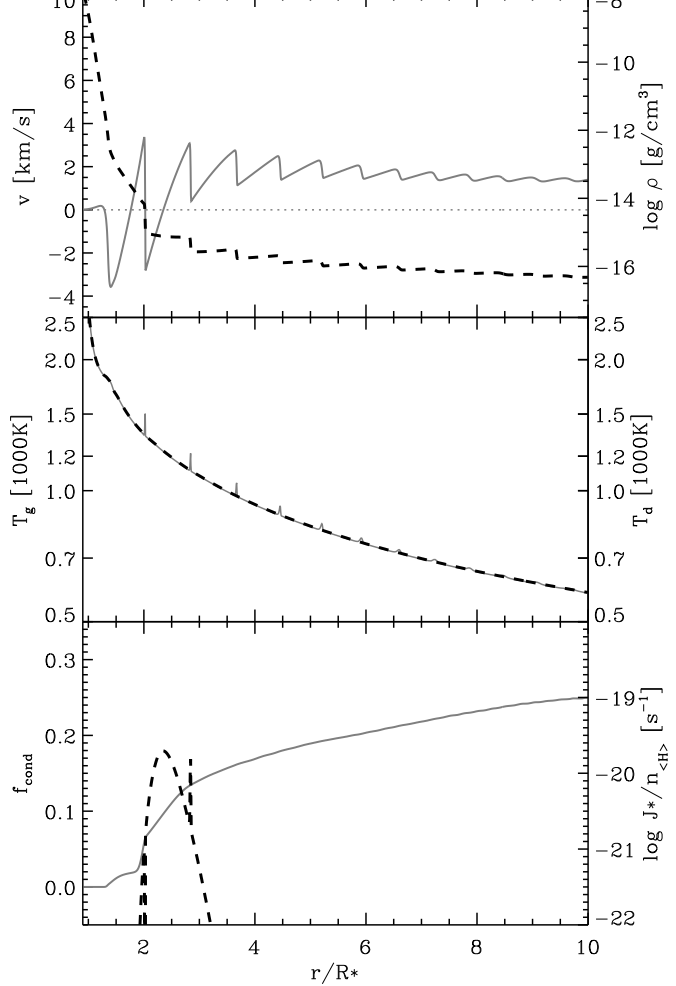, width=8.8cm}
  \vspace*{-5mm}
  \caption{Subsonic 1D solution with small oscillations for
           $M_\star=1\,M_\odot$, $L_\star=10^4L_\odot$, $T_{\rm
           eff}=2600\,$K obtained by {\it decreasing} $\rm
           C/O\,{\scriptstyle\searrow}\,1.6$ after about 1800 years of
           simulation.  Note the small but non-zero degree of
           condensation inside the nucleation peak. For clarity, we
           have ``frozen'' this solution lately by not calling the MC
           radiative transfer routine anymore (see
           Fig.~\ref{fig:1D_stationaer} and text).}
  \label{fig:1D_breeze}
  \vspace*{-2mm}
\end{figure}

The reason for the folding bifurcation sketched in Fig.~\ref{fig:bifu}
lies in an accumulation of small amounts of dust particles in the
``metastable'' region close to the star which is too hot for
nucleation, but too cold for dust evaporation. In a stationary wind,
this region remains completely dust-free because of the throughout
positive radial velocities. However, in case of oscillations, some
dust particles (which have been created farther out) periodically
re-enter this region, where they temporarily stay and grow up to very
big particles with radii as large as $\langle
a\rangle\!=\!\big(\frac{3}{4\pi}\big)^{1/3} L_1/L_0 \approx
50\,\mu$m. The total surface of these few big grains remains so small,
however, that an average molecule needs about $\tau_{\rm
coll}=\big(v_{\rm th} (36\pi)^{1/3} \rho L_2\big)^{-1} \approx
100\,$yrs to collide with them. Hence, the degree of condensation and
the dust opacity are practically frozen in and remain too small to
cause a net outward acceleration $|a_{\rm rad}|\la|a_{\rm
grav}|$. However, $a_{\rm rad}$ is just large enough to change the
pressure gradient in this quasi-static region, which leads to higher
densities in the outer parts.

This density levitation facilitates dust formation in the outer parts
and leads to a persistent, oscillatory stellar outflow with low
(sometimes subsonic) velocities and a smooth dust distribution, see
Fig.~\ref{fig:1D_breeze}. Winters\etal (2002)\nocite{wbnoj2002}
described similar low-velocity, almost quasi-static (``B-type'') 1D
wind solutions in a certain range of stellar parameters under the
influence of a strong stellar pulsation as inner boundary
condition. Here, we find a same type of wind solution even without
stellar pulsation.

The question remains open whether or not such low-velocity outflows
exist in reality. Jura\etal (2002)\nocite{jura2002} noted that the
close-by semi-regular variable L$_2$\,Pup (spectral type
M5\,IIIe--M6\,IIIe) has in fact a very low-velocity outflow
($v_\infty\!\approx\!1.9\,$km/s). Based on continuum observations,
Jura\etal (2002) estimated the mass-loss rate of L$_2$\,Pup to be
$\dot{M}\!\approx\!3.5\times 10^{-7}\rm\,M_\odot/yr$, whereas based on
molecular line observations, Winters\etal (2002) estimated
$\dot{M}\!\approx\!2\times 10^{-9}\rm\,M_\odot/yr$. Despite this large
scatter, these wind properties seem to match the characteristics of
the low-velocity outflows as found by Winters\etal (2002) and in this
work. However, L$_2$\,Pup is an oxygen-rich star with lower
luminosity, higher stellar mass and higher effective temperature as
assumed in these models, so this link between observations and
theoretical models is thin.

%====================================
\subsection{Results of the 2D models}
%====================================
\label{sec:results_2D}

\begin{figure}
  \hspace*{-2mm}
  \begin{minipage}{8.8cm}
  \epsfig{file=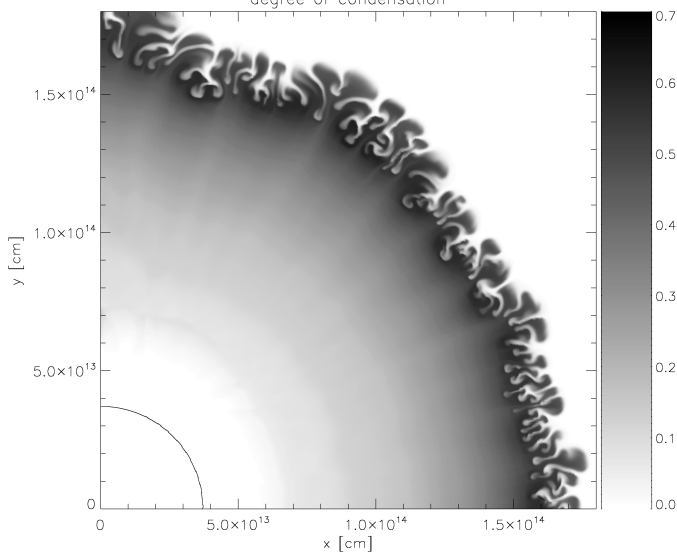, height=8.9cm, angle=90}
  \end{minipage}
  \vspace*{-2mm}
  \caption{Rayleigh-Taylor instabilities in an expanding dust shell
           shortly after the start ($t\!=\!3.2\,$yrs) of a new
           axisymmetric (2D) simulation. Stellar parameters
           $M_\star=1\,M_\odot$, $L_\star=10^4L_\odot$, $T_{\rm
           eff}\!=2500\,$K and $\rm C/O\!=\!1.9$. The additional
           contour line for $\Trad\!=\!2500\,$K indicates the size of
           the star.}
  \label{fig:Rayleigh_Taylor} 
  \vspace*{-1mm}
\end{figure}

The major aim of this work is to advance from 1D to 2D
dust-driven wind models. This enables us, for the first time, to
explore the effects of deviations from spherical symmetry on
dust-driven winds in a consistent way\footnote{Freytag\plus H{\"o}fner
(2003)\nocite{fh2003} have passively integrated the dust moment
equations in first 3D simulations, \ie without the important feedbacks
of the dust on the acceleration and on the radiative transfer.}.  Can
these complicate physico-chemical systems, which have been proven to
be unstable already in 1D models, keep their spherical symmetry, or
--- do instabilities lead to spontaneous symmetry breaking, structure
formation and more complex flow patterns, for example turbulent
mixing? The answer of the 2D simulations is quite unambiguous.

\subsubsection{The Rayleigh-Taylor type instability}
%===================================================
\label{sec:RayleighTaylor}
Already shortly after the start of a new simulation, the expanding
dust shells experience Rayleigh-Taylor type instabilities at their
outer edges (see Fig.~\ref{fig:Rayleigh_Taylor}), which break the
spherical symmetry spontaneously. This instability will be further
analysed and discussed in a forthcoming paper (Woitke, Farzinnia\plus
Weterings 2006)\nocite{wfw2006}, but their nature is easy to
understand. The dust containing gas, which has a larger opacity
$\kext$ and hence a higher potential energy $-G M_\star(1-\alpha)/r$
with $\alpha\!=\!\kext L_\star/(4\pi c GM_\star)$ (see
Eq.~\ref{eq:alpha}), is accelerated into a dust-poor gas with lower
opacity and hence lower potential energy. Thus, it is energetically
favourable to exchange matter across the outer interface of a dust
shell, which makes the fluid unstable.

According to the 2D simulations, a spontaneous mixing occurs in most cases
at the outer edges of dust shells, where dust-rich fingers penetrate
outward and more transparent dust-poor fingers penetrate inward with
respect to the outward moving shell. The occurrence of
Rayleigh-Taylor type instabilities is a persistent feature also in the
more relaxed 2D models.

\begin{figure*}
  \begin{tabular}{cc}
  \hspace*{-3mm}
  \begin{minipage}{9cm}
  \epsfig{file=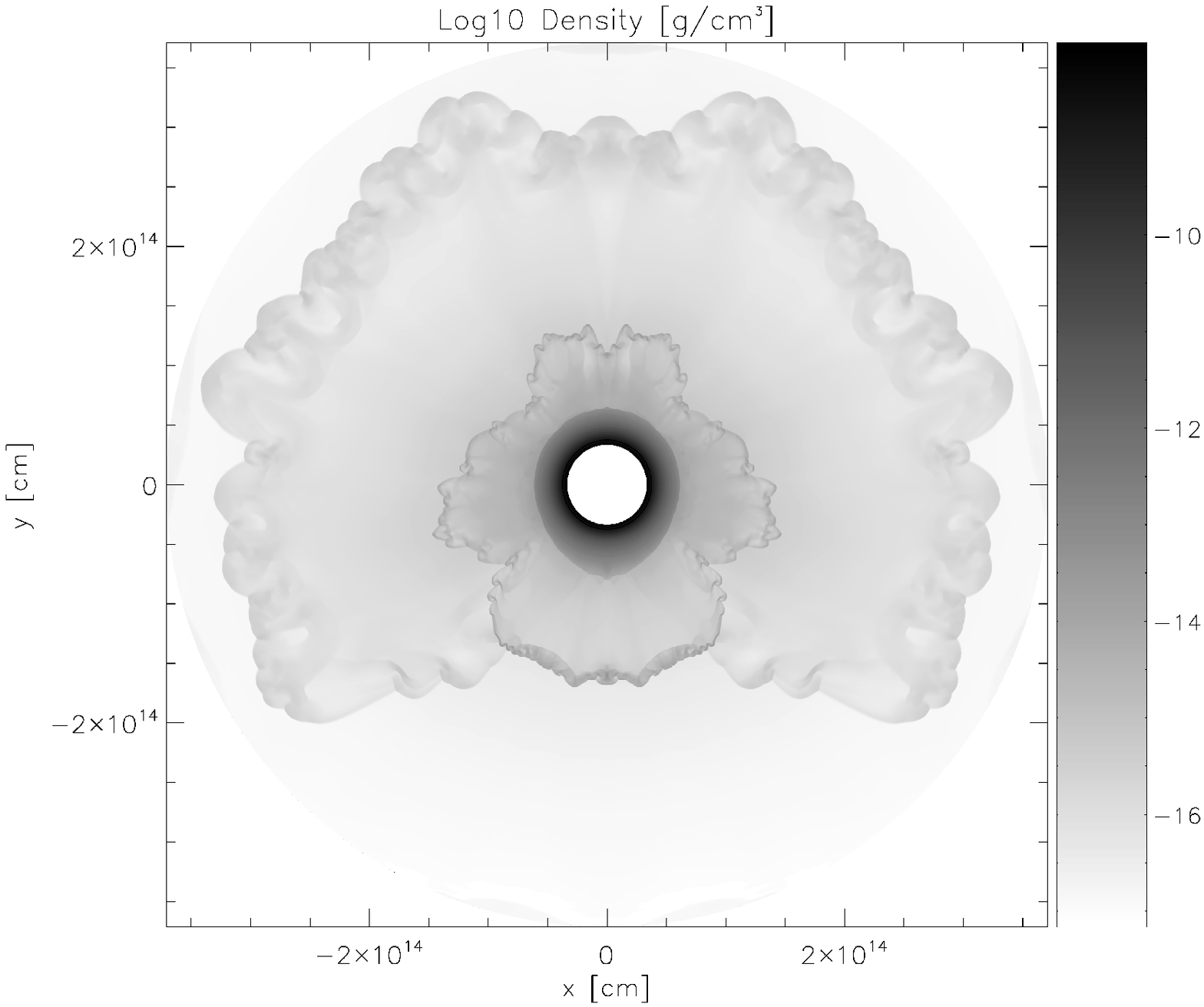, height=9.1cm, angle=90}
  \end{minipage} &
  \hspace*{-6mm}
  \begin{minipage}{9cm}
  \epsfig{file=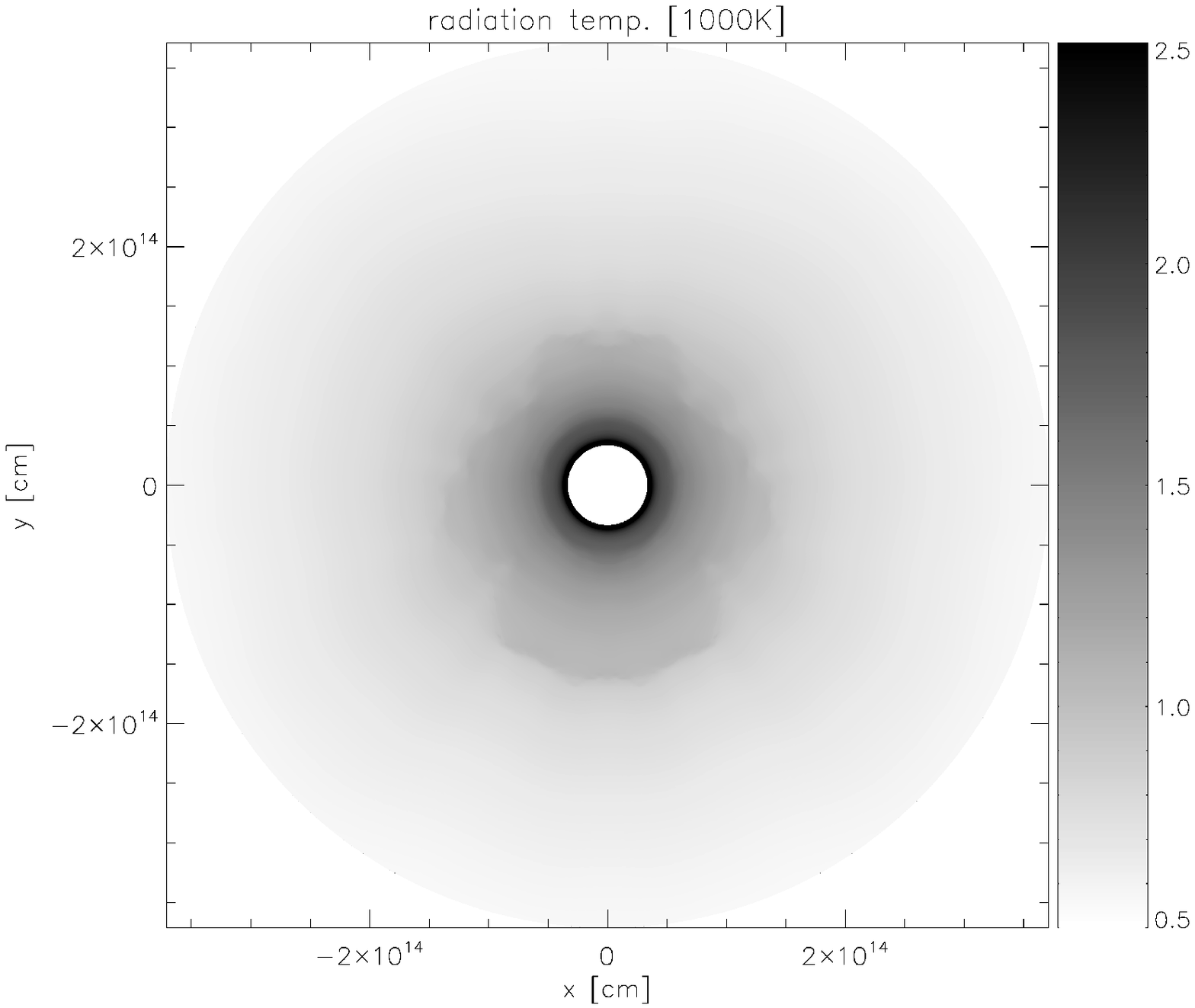, height=9.1cm, angle=90}
  \end{minipage} \\
  \hspace*{-3mm}
  \begin{minipage}{9cm}
  \epsfig{file=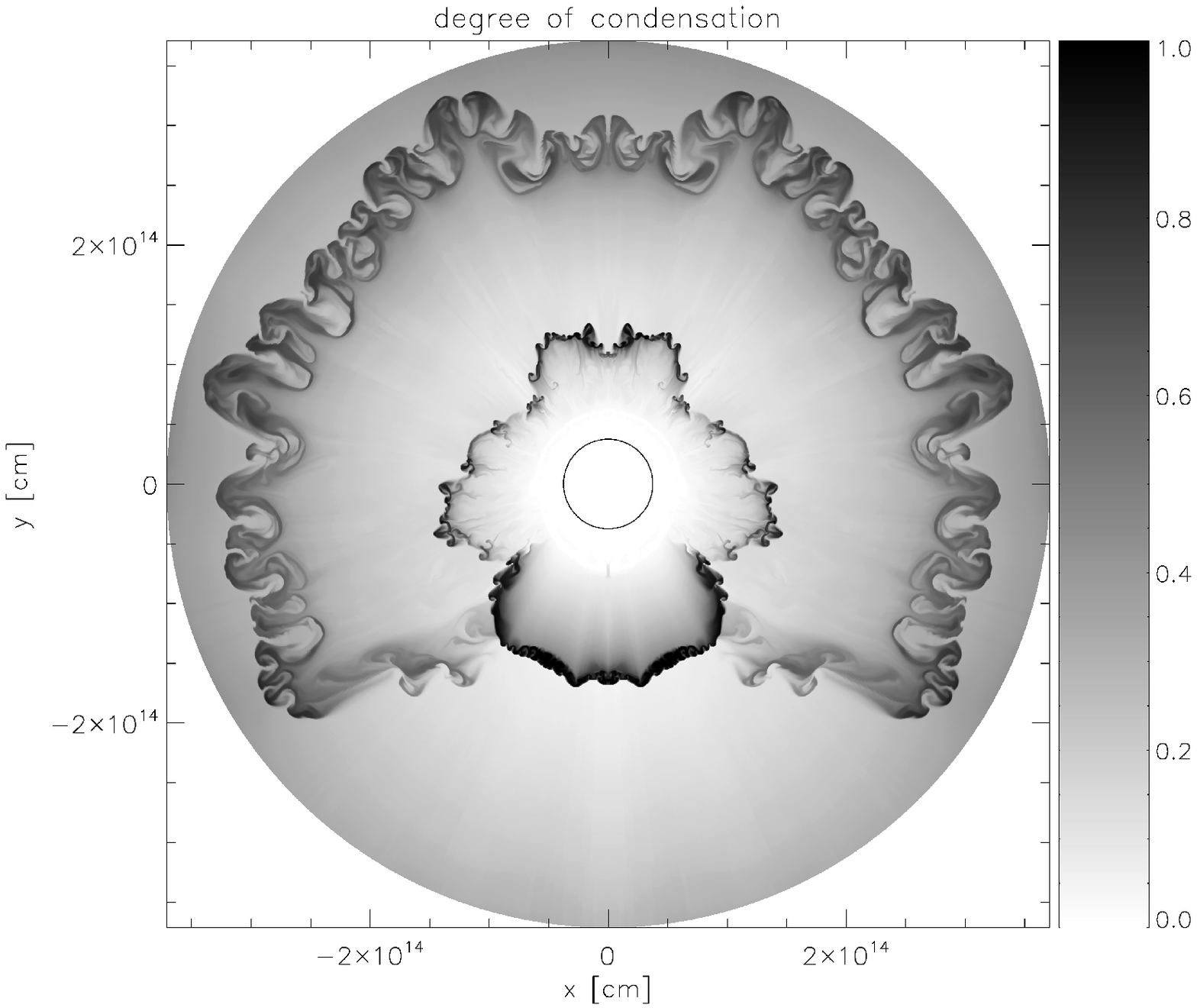, height=9.1cm, angle=90}
  \end{minipage} &
  \hspace*{-6mm}
  \begin{minipage}{9cm}
  \epsfig{file=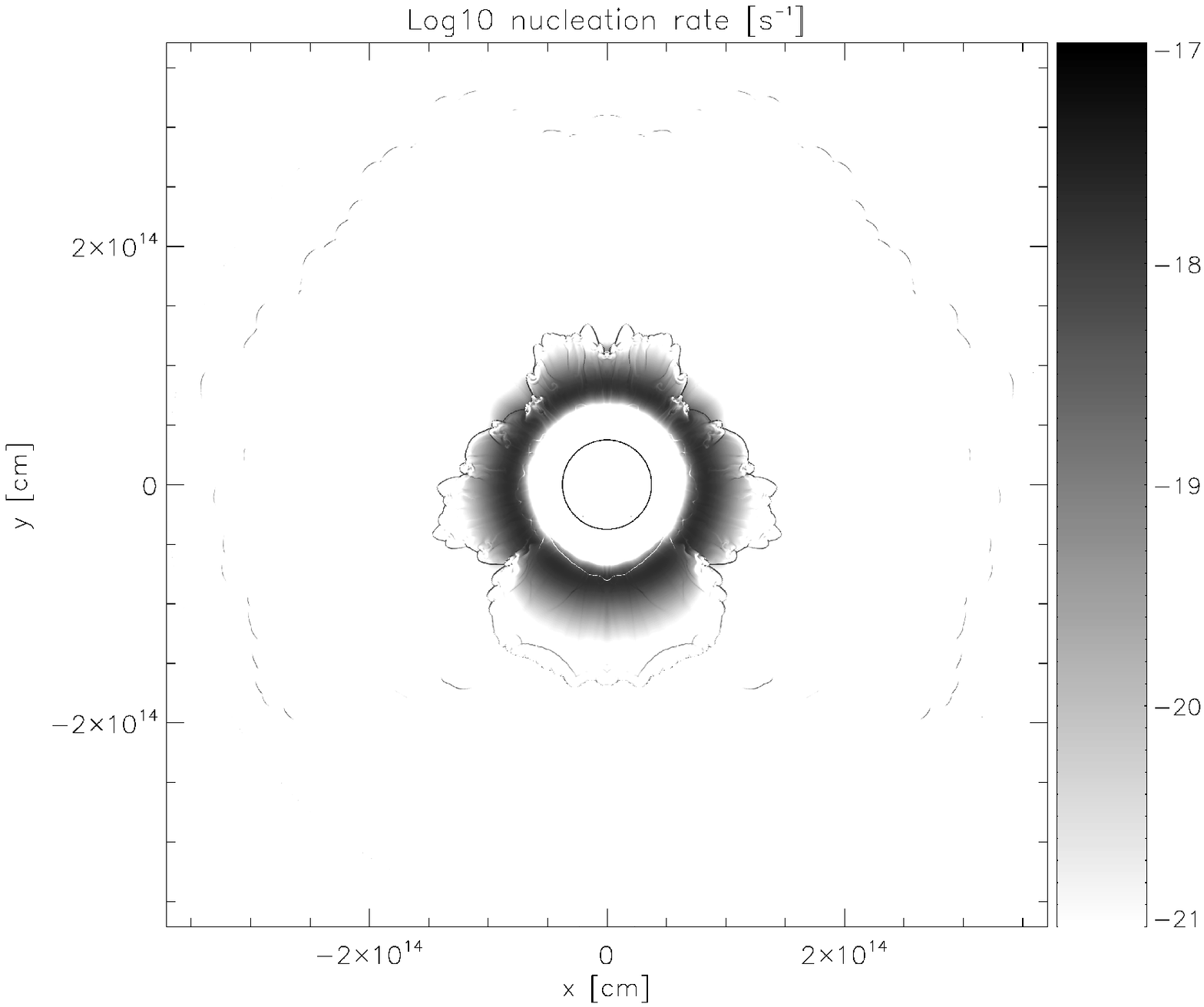, height=9.1cm, angle=90}
  \end{minipage} \\
  \hspace*{-3mm}
  \begin{minipage}{9cm}
  \epsfig{file=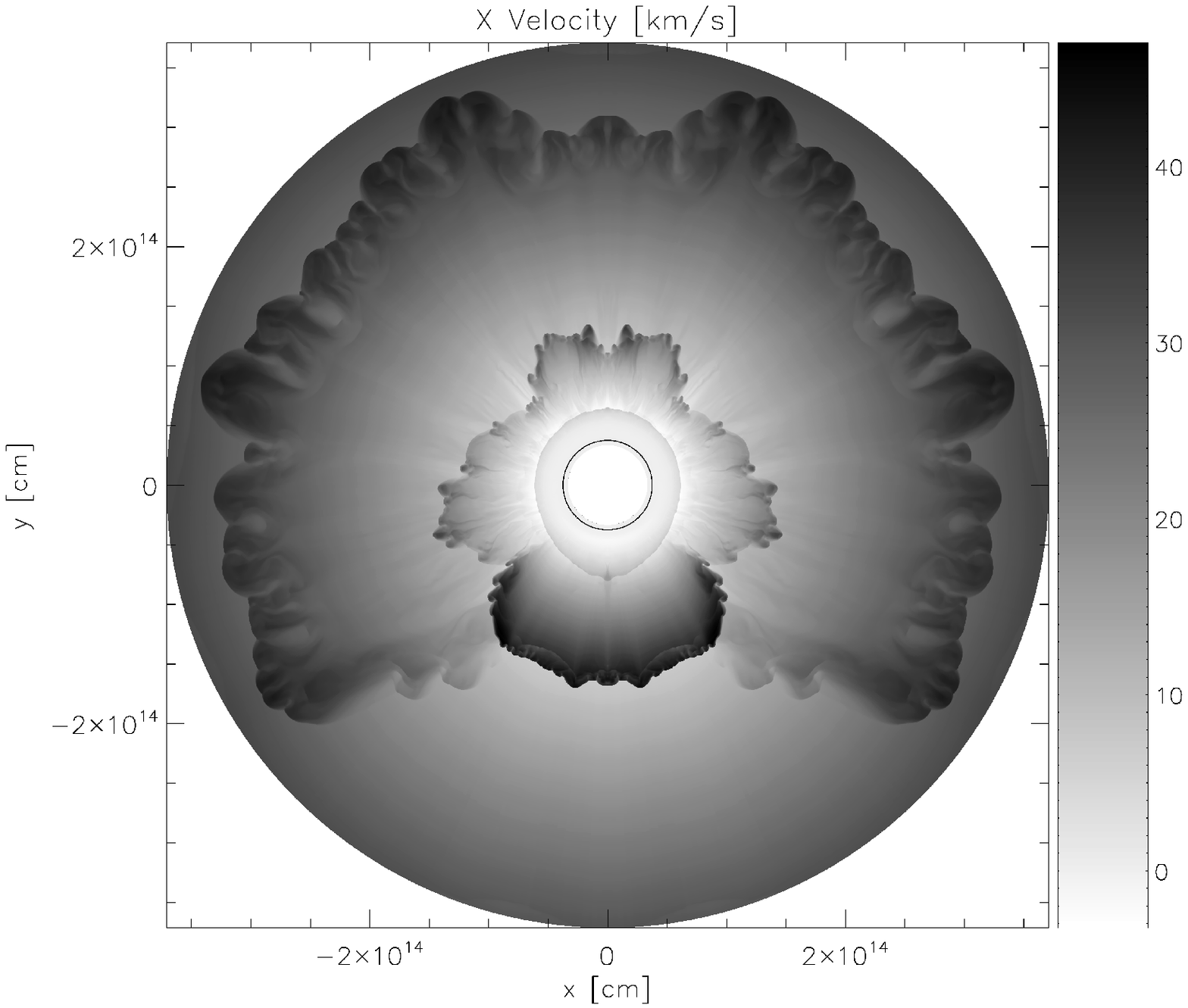, height=9.1cm, angle=90}
  \end{minipage} &
  \hspace*{-6mm}
  \begin{minipage}{9cm}
  \epsfig{file=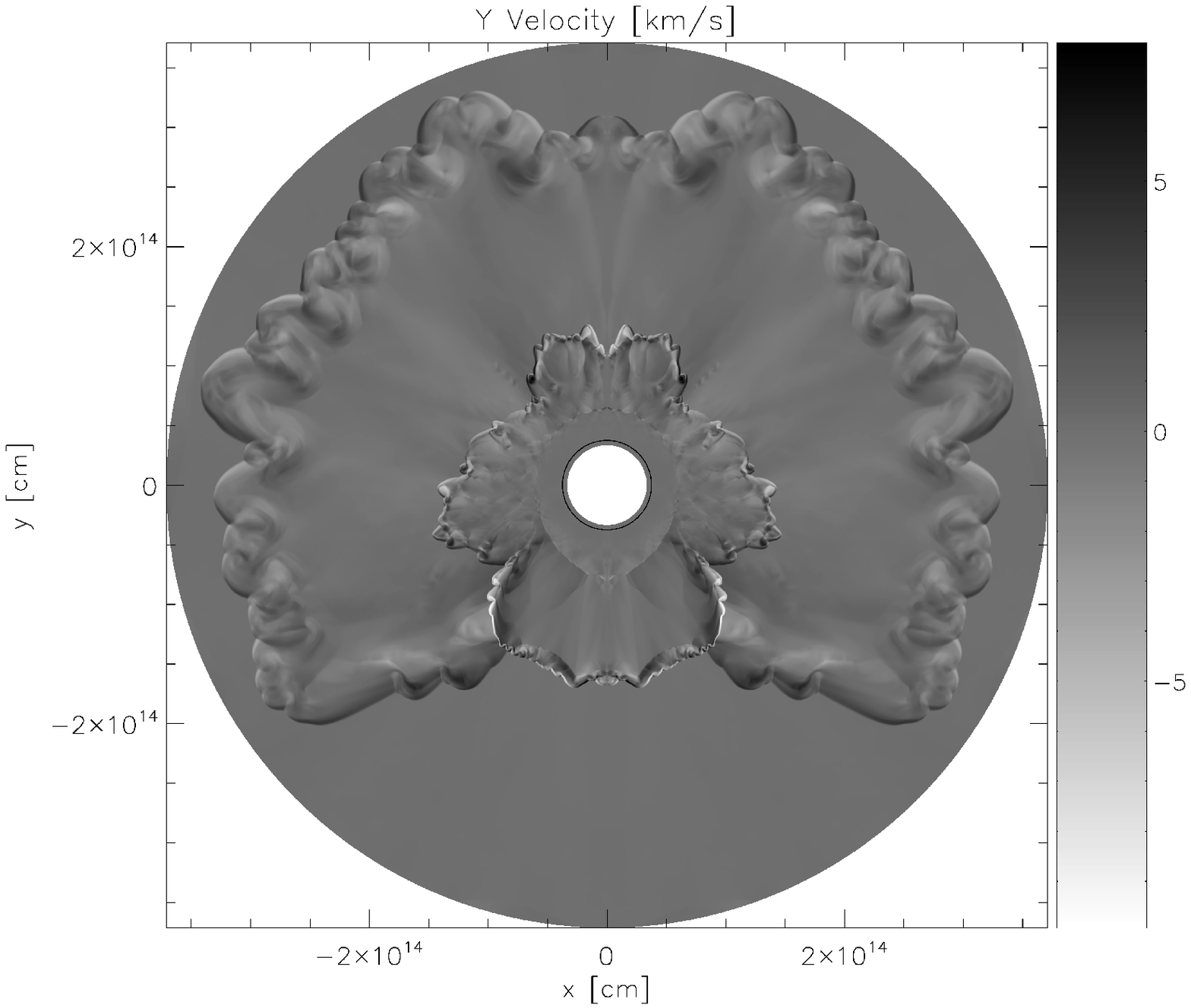, height=9.1cm, angle=90}
  \end{minipage}
  \end{tabular}
  \caption{Snap-shots after 46.5 years of simulation from an
    axisymmetric (2D) model with parameters $M_\star=1\,M_\odot$,
    $L_\star=10^4L_\odot$, $T_{\rm eff}\!=2500\,$K and $\rm
    C/O\!=\!2.2$, showing the following quantities as grey-scale
    contour plots: {\bf first row:} mass density $log\,\rho$ and
    radiation temperature $\Trad$, {\bf second row:} degree of
    condensation $f_{\rm cond}$ and nucleation rate per hydrogen
    nucleus $\log(J_\star/\nH)$, {\bf last row:} radial velocity $v_r$
    and tangential velocity $v_\theta$ (white indicates a motion
    towards the north pole and black towards the south pole). The
    additional black contour line for $\Trad\!=\!2500\,$K indicates
    the size of the star. Central blank regions are not included in
    the model.}
\label{fig:2D_model}
\end{figure*}

%==========================================
\subsubsection{Description of 2D solutions}
%==========================================
\label{sec:2D}

In the following, we will describe the qualitative behaviour of two
relaxed 2D models with emphasis on the main new features caused by
dropping the assumption of spherical symmetry\footnote{Unfortunately,
the 2D models are so expensive (see Sect.~\ref{sec:limits}) that only
two complete high-resolution 2D simulations are available up to now,
supplemented by a number of additional low-resolution
simulations.}.
%General conclusions about the dependence of the 2D
%results on the stellar parameters are not possible at this point.}.

The first high-resolution 2D model has been calculated for parameters
$M_\star=1\,M_\odot$, $L_\star=10^4L_\odot$, $T_{\rm eff}\!=2500\,$K
and $\rm C/O\!=\!1.9$.  This set of parameters results in an optically
thin stationary wind with only little $\theta$-variations. The radial
structures (cuts along constant $\theta$) resembles well the
respective 1D model structure (see Sect.~\ref{sec:stationary_winds}).

The second high-resolution 2D model has been calculated for the same
stellar parameters, but now with $\rm C/O\!=\!2.2$, a parameter
combination which resulted in an optically thick wind with irregular
oscillating behaviour in 1D (see Table~\ref{tab1}).
Figure~\ref{fig:2D_model} shows some details of this 2D model after
$46.5\,$yrs of simulation, after which the respective 1D model is
sufficiently relaxed.

The most striking new feature is that the dust ``shells'' are not
complete, but do only occupy a certain fraction of the total solid
angle.  These dust ``arcs'', sometimes also smaller ``caps'' originate
from the dust formation zone which is mainly radially oscillating
according to the exterior $\kappa$-mechanism known from the 1D models
(see Sect.~\ref{sec:osc}).  However, these oscillation often become
out-of-phase concerning north-polar, equatorial and south-polar
regions, etc., in the final relaxed state. Consequently, dust forms
from time to time in restricted regions above the stellar surface.
Only in those cases where the complete dust formation zone around the
star performs a concerted radial oscillation, a complete radial dust
shell is produced. In all other cases, the production of limited dust
arcs is more typical.

In fact, the different polar and equatorial regions seem to perform
more or less their own independent oscillations, since
they are only weakly coupled to each other via tangential velocities
and radiative transfer feedbacks. Considering the often slightly
chaotic behaviour of the dust formation zone in the 1D models
(weak chaos), it's clear that small disturbances (\eg introduced by
the Monte Carlo noise or by the Rayleigh-Taylor instability) may cause
completely different trajectories after sufficiently long times.

There is, however, one mechanism that tends to equilibrate the phases
of the oscillations and to reinstall spherical symmetry, namely the
radiative backwarming, which interrupts the dust production inside a
forming shell (see Sect.~\ref{sec:osc}). However, this backwarming is
not as simple as in 1D geometry. It has only a limited reach and
cannot trigger the dust formation at the opposite side of the star. We
believe that this is the reason why the typical angular extent of the
dust arcs in the model is about one forth to one half of the total
solid angle: An arc of such an angular extension produces locally
almost as much backwarming as a complete radial shell.

\begin{figure}
  \hspace*{-2mm}
  \begin{minipage}{8.8cm}
  \epsfig{file=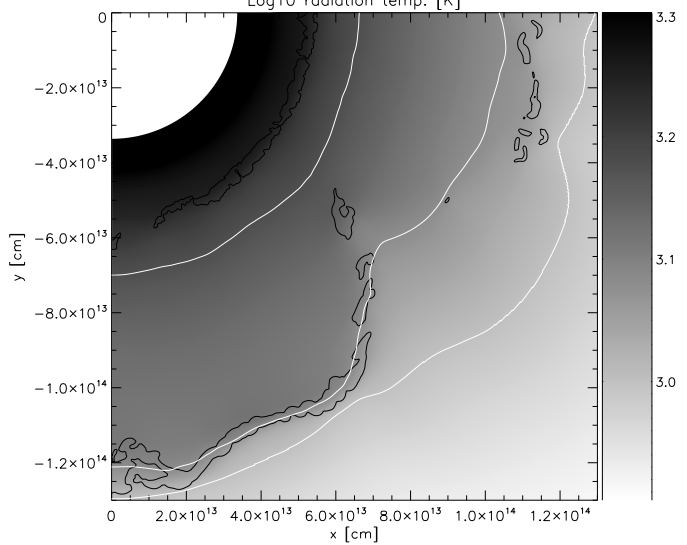, height=8.9cm, angle=90}
  \end{minipage}
  \vspace*{-2mm}
  \caption{Details of the radiation temperature structure in the 2D
           model shown in Fig.~\ref{fig:2D_model} (same timestep) with
           overplotted white contour lines for $\Trad\!=\!1000\,{\rm K}, 
           1200\,$K and 1500\,K. The additional black contour line for 
           $\kappa_{\rm ext}\!=\!5\times 10^{-14}\rm\,cm^{-1}$ encircles the
           optically thick dust structures. Note the backwarming of
           the dust arc at the bottom and the formation of shadows
           behind the opaque regions.}
  \label{fig:Trad_details} 
  \vspace*{-1mm}
\end{figure}

Figure~\ref{fig:Trad_details} shows further details of the calculated
temperature structures. The radiative transfer through
the clumpy dust arcs causes complex temperature patterns with
backwarming, sidewarming and shadow formation. The consequence is
that the next dust formation episode will occur preferentially
in those regions close to the star which are now not so much affected by
the radiative backwarming.

Once a dust arc has formed, it is accelerated outward by radiation
pressure. This outward motion causes not only a radial displacement
and dilution, but also a tangential expansion because of the
decreasing ambient pressure. Supersonic tangential velocities result
in this way (see lower r.h.s. part of Fig.~\ref{fig:2D_model}) which
increase the angular size of the arcs. At the side edges of the dust
arcs, the matter stays behind and continues to form more dust which
leads to an apparent inward bending and extension of the arcs.
Concerning the smaller dust caps, the expansion motion resembles the
behaviour of mushroom clouds: tangential expansion at the top and
counter-rotating vortices in the wake with Kelvin-Helmholtz
instabilities.

The occurrence of negative (infall) velocities underneath and within
the dust formation zone is a typical features in all oscillating wind
models, not only in 1D but also in 2D. However, in the 2D case,
dust-poor gas may fall back toward the star at the same time as
dust-rich gas is accelerated outward. In fact, these infall motions
occur typically just between the outward moving arcs and clouds. Since
the infalling gas is compressed by the ambient pressure, the infalling
dust-poor regions are usually much smaller than the outward moving
dust-rich structures.

The sudden radiative acceleration of the newly formed dusty regions
leaves behind temporary gaps in the dust formation zone which are not
only refilled by dust-free matter streaming in from the stellar
atmosphere, but also by infalling matter falling back from the
outside. These two streams collide in the dust formation zone, which
creates turbulence. This turbulence provides additional reasons for
the subsequent formation of inhomogeneous dust distributions
(Helling\etal 2004)\nocite{hkwns2004}.

Furthermore, the infalling matter may already contain some small
amounts of dust particles which have been created by nucleation
farther out. In this way, dust particles reach again the much
denser dust formation zone, where they will serve as additional
nucleation seeds. This process results to be the main cause for new
dust formation events. Considering the history of sample grains, we
find typically

\smallskip
\centerline{nucleation $\to$ infall $\to$ growth $\to$ acceleration.}
\smallskip

\noindent Last but not least, the more distant dust shells and arcs
exhibit a variety of small-scale sub-structures. These small-scale
structures are the result of the action of various flow instabilities,
not only the Rayleigh-Taylor instability as briefly explained in
Sect.~\ref{sec:RayleighTaylor}.  These instabilities apparently need
some time to amplify perturbations and so to further shape and
fragment the outward moving dust shells and arcs. The result are
cauliflower-like shapes, which could also be described as numerous
small-scale clouds within the expanding dust shells and arcs.

%===========================================
\subsubsection{Comparing 1D with 2D results}
%===========================================

\begin{table}
\caption{Comparison of calculated wind quantities between 1D and 2D models
         for $M_\star=1\,M_\odot$, $L_\star=10^4L_\odot$, 
         $T_{\rm eff}=2500\,$K and $\rm C/O=2.2$.}
\label{tab3}
\begin{center}
\begin{tabular}{l|c|c|c}
       geometry
       & $\langle\dot{M}\rangle[M_\odot/{\rm yr}]$ 
       & $\langle v_\infty\rangle[\rm km/s]$
       & $\langle\rho_{\rm d}/\rho_{\rm g}\rangle$\\
&&&\\[-2.5ex]
\hline
  1D (spheric.~sym.) & $7.9(-6)$ &  31.7  & $3.2(-3)$\\
  2D (axisymmetric)  & $5.6(-6)$ &  29.2  & $3.2(-3)$
\end{tabular}
\end{center}
\end{table}

Considering the calculated mean wind quantities (\eg the mass loss
rate), the results of the 1D and 2D models are remarkably similar (see
Table~\ref{tab3}), despite all the detailed effects described in
Sect.~\ref{sec:2D}. The slightly lower mass loss rate in the 2D model
(30\%) could indicate that dust arcs are slighly less effective in
lifting the dust-poor gas in between out of the gravitational
potential of the star than complete dust shells. However, this
deviation could also be a simple consequence of the insufficient time
given to the 2D model to relax.

The deviations along one specific line of sight, however, may be quite
different. Consequently, the spectral appearance of the star can be
strongly affected by the deviations from spherical symmetry
presented in this paper. This will concern lightcurves, images and
visibilities, in particular at short (optical and near infrared)
wavelengths, where the flux consists mainly of dust attenuated stellar
photons. These effects need to be further investigated in a
forthcoming paper.

%----------------------------------------------------------------------------
\section{Summary and discussion}
%----------------------------------------------------------------------------

New axisymmetric (2D) hydrodynamical models for carbon-rich AGB star
winds have been developed which include coupled grey continuum Monte
Carlo radiative transfer, equilibrium chemistry, time-dependent dust
formation and radiation pressure on dust. According to our knowledge,
this is actually the first successful application of the Monte Carlo
radiative transfer technique in the frame of multi-dimensional
hydrodynamical simulations.

The spherical symmetric (1D) version of the code has been checked
against other published works (Fleischer et al.\ 1995, H{\"o}fner\etal
1995) showing good agreement in qualitative wind behaviour and
quantitative results. Considering the most simple case without stellar
pulsation (hydrostatic inner boundary condition), the 1D simulations
reveal 3 basic types of dust-driven stellar winds: {\it stationary
winds} with low mass-loss rates and low but supersonic outflow
velocities, {\it oscillating winds} with shock waves and dust shells,
high mass-loss rates and high outflow velocities, and {\it oscillating
low-velocity outflows} with low mass-loss rates and sometimes even
subsonic outflow velocities.

All three types of dust-driven winds may occur for the same stellar
parameters $M_\star$, $L_\star$ and $T_{\rm eff}$ if the
carbon-to-oxygen ratio C/O is considered as free parameter. The
dependence of wind type (and resultant wind properties) on C/O is
found to be {\it non-unique}, \ie dependent on history. By increasing C/O,
we find a sudden jump from stationary to oscillating wind solutions,
whereas for decreasing C/O, we find a sudden jump from oscillating
winds to oscillating low-velocity outflows.  This {\it
folding bifurcation} (hysteresis behaviour) has not been reported so
far. The third type of wind solution (the oscillating low-velocity
outflows) seem to resemble the B-type wind solutions proposed by
Winters\etal (2002), although we have not considered any stellar
pulsation in this paper.

The possibility of a star with given parameters $M_\star$, $L_\star$,
$T_{\rm eff}$ and C/O to produce two different types of winds with
high and low mass-loss rates might open up the possibility of
spontaneous or induced switches between these two mass-loss modes,
which could be related to the observation of distant radial dust
shells (see \eg Mauron \plus Huggins 1999). A mechanism to obtain such
switches has been reported by Simis\etal (2001).\nocite{mh99,sid2001}

The results of the new axisymmetric (2D) dust-driven wind models
reveals an even more complex picture of the dust and wind formation
around carbon-rich AGB stars. Although the calculated mean
wind properties like $\langle\dot{M}\rangle$ and $\langle v_\infty\rangle$
of the new 2D wind models are rather similar to those of the 
1D models, the details exhibit several new effects.

Despite the spherically symmetric initial and boundary conditions,
spontaneous symmetry breaking occurs in these winds due to action of
various instabilities. Dust is found to form from time to time in
restricted regions above the stellar surface. These spatially
restricted dust formation events lead to the production of incomplete
dust shells, dust arcs or even smaller dust caps. Only in those cases
where the complete dust formation zone around the star performs a
concerted radial oscillation, a complete dust shell is produced.

The opaque but geometrically thin dust structures are accelerated
outward by radiation pressure, expanding radially and tangentially
like mushroom clouds, which leads to an accumulation of overlapping arcs
in the more distant regions. The optically thin, dust-poor matter in
between tends to fall back toward the star, not only temporal (between
the active dust formation phases) but also spatial (between the
outward moving dusty structures). This matter mixes with fresh
dust-free matter from the stellar atmosphere in a turbulent way in
the dust formation zone, which again leads to a subsequent production of
irregular dust structures.

Further away from the star, flow instabilities (\eg Rayleigh-Taylor)
have time to modify and fragment the outward moving dust arcs and
shells, producing numerous small-scale clumps in the outward
moving dust shells and arcs.

From the few models calculated so far, we conclude that whenever the
1D model relaxes to an irregular oscillatory wind solution, the 2D
model is likely to show strong deviations from spherical symmetry.
This type of solution is typical for optically thick winds of embedded
IR carbon stars.  In contrast, if the 1D model relaxes toward a
stationary wind solution (which is the typical result for optical carbon 
stars with optically thin winds), the 2D model shows a similar behaviour.

The formation of incomplete shells, arcs or smaller caps due to an
irregular dust production of the central star may be important to
understand recent interferometric observations of embedded carbon
stars and red supergiants (\eg Monnier\etal 2004), which show obvious
deviations from spherical symmetry. Woitke\plus Quirrenbach (2005)
have calculated IR images and visibilities from the
calculated 2D models discussed in this paper\footnote{See
{\tt http://www.strw.leidenuniv.nl/$\sim$woitke} for mpeg-animations and 
simulated images.} which can be useful for the interpretation
of such observations.

\begin{acknowledgements}
  I want to express my gratefulness to the Leiden theory group around
  V.~Icke for the support of my work. In particular, I want to
  thank E.-J.~Rijkhorst for introducing me to the {\sc Flash}-code
  and S.-J.~Paardekooper for correcting the manuscript.
  This work is part of the {\sc AstroHydro3D} initiative supported by
  the {\sc NWO Computational Physics programme}, grant
  614.031.017. The software used in this work was in part developed by
  the DOE-supported ASCI/Alliance Center for Astrophysical
  Thermonuclear Flashes at the University of Chicago.
\end{acknowledgements}

%\begin{appendix}
%\end{appendix}

%\bibliographystyle{aa}
%\bibliography{../ref801}

\end{document}